\documentclass[preprint]{aastex}
\usepackage{emulateapj5}

\shorttitle{Excess Entropy and the SZ Effect}
\shortauthors{McCarthy et al.} 

\begin{document} 

\title{The SZ Effect Signature of Excess Entropy in Distant, Massive Clusters}

\author{Ian G. McCarthy$^{1,2}$, Gilbert P. Holder$^3$, Arif Babul$^{1,4}$, and Michael L. 
Balogh$^5$}

\affil{$^1$Department of Physics \& Astronomy, University of Victoria, Victoria, BC, 
V8P 1A1, Canada}

\affil{$^3$School of Natural Sciences, Institute for Advanced Study, Princeton, NJ, 
08540, USA}

\affil{$^5$Department of Physics, University of Durham, South Road, Durham, DH1 3LE, 
UK}

\footnotetext[2]{email: mccarthy@beluga.phys.uvic.ca}
\footnotetext[4]{CITA Senior Fellow}

\begin{abstract} 

Studies of cluster X-ray scaling relations have led to suggestions that non-gravitational 
processes, e.g., radiative cooling and/or ``preheating'', have significantly modified the 
entropy of the intracluster medium (ICM).  For the first time, we test this hypothesis 
through a comparison of predicted thermal Sunyaev-Zeldovich (SZ) effect scaling relations 
with available data from the literature.  One of the relations that we explore, in 
principle, depends solely on SZ effect observations, thus offering an X-ray independent 
probe of the ICM.  A detailed comparison of the theoretical relations with the largest 
compilation of high-$z$ SZ effect data to date indicates that the presence of an entropy 
floor is favored by the data.  Furthermore, the inferred level of that floor, $K_0 
\gtrsim 300$ keV cm$^2$, is comparable to that found in studies of X-ray scaling 
relations of nearby massive clusters.  Thus, we find no evidence for significant 
evolution of the entropy floor out to $z \sim 0.7$.  We further demonstrate that the high 
quality data to be obtained from the upcoming Sunyaev-Zeldovich Array ({\it SZA}) and the 
(soon-to-be) upgraded Owens Valley Radio Observatory ({\it OVRO}) array will open 
powerful new windows into the properties of the ICM.  Specifically, the new measurements 
will allow for accurate measurements of the ICM entropy for even the most distant galaxy 
clusters.

\end{abstract}


\section{INTRODUCTION}

The failure of theoretical self-similar X-ray scaling relations to match observed trends 
has led to suggestions that important non-gravitational processes, such as radiative 
cooling and/or ``preheating'', are significantly affecting the structure and appearance of 
the intracluster medium (ICM).  Models and simulations of clusters which attempt to 
explicitly take into account the effects of cooling and/or heating produce clusters which 
have higher mean entropies (``excess'' entropy) than those produced by models which neglect 
these processes.  In some cases, the model clusters possess cores in their entropy profiles 
commonly referred to as the ``entropy floor''.  The presence of this entropy floor, in 
turn, modifies the predicted X-ray scaling relations, bringing them in to much closer 
agreement with the observed correlations (e.g., Kaiser 1991; Evrard \& Henry 1991; Bower 
1997; Balogh et al. 1999; Wu et al. 2000; Bryan 2000; Tozzi \& Norman 2001; Borgani et al. 
2001; Voit \& Bryan 2001; Babul et al. 2002; McCarthy et al. 2002; Thomas et al. 2002; 
Voit et al. 2002; Dav\'{e} et al. 2002; Lloyd-Davies et al. 2003).  Direct observational 
evidence for an ``entropy floor'' in nearby groups and low mass clusters has been 
presented by Ponman et al. (1999) and Lloyd-Davies et al. (2000).

To date, only X-ray observations have yielded information about the entropy floor but 
because the X-ray surface brightnesses of groups/clusters suffer the effects of cosmological 
dimming [the bolometric surface brightness scales as $(1+z)^{-4}$], both direct and 
indirect studies of the entropy floor via X-ray observations have generally been limited to 
low redshift ($z \lesssim 0.2$) systems.  An {\it independent} test of the entropy floor 
hypothesis which could also provide information on {\it high redshift} clusters and, 
therefore, the evolution of the non-gravitational processes that give rise to the entropy 
floor, would be extremely useful.  In a companion paper (McCarthy et al. 2003, 
hereafter MBHB03), we argued that a number of scaling relations based entirely or in 
part on thermal Sunyaev-Zeldovich effect (Sunyaev \& Zeldovich 1972; 1980; hereafter, 
referred to as the SZ effect) observables can, potentially, be used for both of these 
purposes.  Our analysis indicated that even current SZ effect observational data from, 
for example, the Berkeley Maryland Illinois Association ({\it BIMA}) and Owens Valley 
Radio Observatory ({\it OVRO}) arrays and the Ryle Telescope, when compared to the 
predicted correlations, can be used to tell us something about the entropy floors of 
distant, massive clusters.  

The primary focus of the present paper is a comparison of these predicted scaling 
relations with available data from the literature to determine if the SZ effect data 
favor the existence of an entropy floor and how the inferred level of the entropy floor 
compares with that required to explain local X-ray trends (which require $K_0 \gtrsim 
300$ keV cm$^2$ for massive clusters; Tozzi \& Norman 2001; Babul et al. 2002; McCarthy 
et al. 2002).  {\it This is the first time that the SZ effect has been used as a probe of 
non-gravitational entropy in galaxy clusters.}  

A number of new SZ experiments, which will greatly improve the quality of the 
observations, are being planned or are already under construction.  An additional goal of 
the present study, therefore, is to examine the efficacy of two of these experiments, the 
Sunyaev-Zeldovich Array ({\it SZA}) and the (soon to be) upgraded {\it OVRO} array, to 
constrain the properties of the excess entropy in distant clusters.  By constructing and 
analysing ``mock'' observations which take explicit account of the expected instrumental 
responses of these arrays, we quantify the accuracy with which the level of the entropy 
floors of distant clusters can be inferred by future data as a function of redshift.

The present paper is outlined as follows: in \S 2, we discuss and select available 
data from the literature; in \S 3, we compare these data to our theoretical scaling 
relations; in \S 4, we assess the ability of the upcoming {\it SZA} and upgraded {\it 
OVRO} array to probe the entropy floors of distant clusters; and in \S 5; we discuss and 
summarize our results.  We assume $\Omega_m = 0.3$, $\Omega_{\Lambda} = 0.7$, and $H_0 = 
75$ km s$^{-1}$ Mpc$^{-1}$ and work in physical units (e.g., $M_{\odot}$ rather than 
$h^{-1} M_{\odot}$) throughout the paper.

\section{Observational data}

Observations of the SZ effect have advanced tremendously over the last decade or so and 
routine high signal-to-noise measurements of the effect are now being made with a 
variety of instruments (including both single-dish and interferometric experiments) at 
a variety of wavelengths (spanning radio down to the submillimeter).  There now exist 
published data for some 30-40 clusters (e.g., Jones et al. 1993; Herbig et al. 1995; 
Carlstrom et al. 1996; Myers et al. 1997; Holzapfel et al. 1997; Hughes \& Birkinshaw 1998; 
Pointecouteau et al. 1999; 2001; 2002; Patel et al. 2000; Holzapfel et al. 2000; Joy et al. 
2001; Mason et al. 2001; Grego et al. 2001; Grainge et al. 2002a; Grainge et al. 2002b; 
Jones et al. 2003; Reese et al. 2002; Cotter et al. 2002; LaRoque et al. 2003; Cantalupo 
et al. 2003).  With this influx of new data, the sample is large enough to make 
statistically significant comparisons between observed and theoretically predicted SZ 
effect scaling relations.  Such comparisons test our understanding of the ICM and 
clusters in general.  In this section, we compile and discuss cluster SZ effect data from 
the literature with the intention of comparing it to theoretically predicted scaling 
relations in \S 3.

Of the 30-40 clusters that have published SZ effect data, we are particularly interested in 
those clusters which lie at high redshift.  As already mentioned, the vast majority of 
studies on the {\it X-ray} scaling relations of clusters have been for low redshift ($z 
\lesssim 0.2$) systems but information on the entropy floors of more distant 
clusters is scant.  Because the SZ effect is not subject to cosmological dimming, SZ effect 
scaling relations potentially offer a way of probing the non-gravitational entropy of even 
the most distant galaxy clusters.

A search of the literature for high redshift clusters with SZ effect data yields 22  
different clusters in the range $0.14 \lesssim z \lesssim 0.78$, many of which were 
observed multiple times.  The clusters are listed in Table 1 along with their redshifts, 
central Compton parameters ($y_0$), frequency-independent integrated SZ effect flux 
densities within the central $1\arcmin$ ($S_{\nu,arc}/f_{\nu}$), and the bibliographic 
references (references in parentheses indicate that the cluster was observed multiple 
times).  We are particularly interested in $y_0$ and $S_{\nu,arc}/f_{\nu}$ 
because a number of scaling relations based on these two quantities are expected to 
be quite sensitive to the entropy floor level of galaxy clusters (see MBHB03).  

Below, we discuss how we extract $y_0$ and $S_{\nu,arc}/f_{\nu}$ from the observational 
data.  At present, it is not possible to directly measure either of these quantities.  
Observations of the SZ effect filter large-scale emission while finite resolution smears 
out small-scale structures.  As discussed in MBHB03, fitting a model (such as the 
isothermal $\beta$ model; Cavaliere \& Fusco-Femiano 1976; 1978) to the SZ effect data 
provides a method for effectively removing these effects and estimating $y_0$ and 
$S_{\nu,arc}/f_{\nu}$, but it should be kept in mind that such quantities are inferred 
and model-dependent.  Provided the smallest angular scale resolved is comparable to the 
typical scale over which the cluster varies, the estimated central Compton parameter will 
be reliable, while inferred flux densities will be suspect when extrapolated beyond the 
filtering scale of the observations.  For current interferometric observations [such as 
those obtained with the {\it BIMA} and {\it OVRO} arrays and the Ryle telescope], the 
highest angular resolution for SZ measurements is typically smaller than the core radius 
of the cluster ($\sim 30\arcsec$) while the large-scale filtering normally becomes 
important on scales larger than about $2\arcmin$.  Therefore, it can be expected that the 
inferred values of $y_0$ and $S_{\nu,arc}/f_{\nu}$ (the flux density within the central 
$1\arcmin$) should be accurate.  

To calculate the total frequency-independent SZ effect flux density within the central 
$1\arcmin$ for each of these clusters, we must reconstruct each of the SZ effect
``surface brightness'' profiles, $y(\theta)$.  The majority of the clusters listed in Table 
1 were modeled using the spherical isothermal $\beta$ model.  In this model, the ICM is 
assumed to be isothermal and has a density distribution described by

\begin{equation}
n_e(r) = n_{e0} \biggl(1 + \frac{r^2}{r_c^2} \biggr)^{- 3 \beta /2}
\end{equation}

\noindent were $n_{e0}$ is the central electron density, $r_c$ is the cluster core radius,
and $\beta$ is the power-law index.  This leads to a SZ effect surface brightness profile

\begin{equation}
y(\theta) = y_0 \biggl(1 + \frac{\theta^2}{\theta_c^2} \biggr)^{1/2 - 3 \beta /2}
\end{equation}

\noindent where $y(\theta)$ is the Compton parameter evaluated at a projected position
$\theta = r/D_a$ from the cluster center and is proportional to the integrated pressure
along the line-of-sight through the cluster.  In addition, $\theta_c = r_c/D_a$ ($D_a$ is 
the angular diameter distance).

The central Compton parameters, $y_0$, are converted from the central SZ effect temperature 
decrements (references are given in col. 5) using the relation

\begin{equation}
\frac{\Delta T_0}{T_{CMB}} = y_0 \biggl(\frac{x}{\tanh{x/2}} - 4 \biggr)
\end{equation}

\noindent where $\Delta T_0$ is the central SZ effect temperature decrement and $x$ = $h
\nu/ k T_{CMB}$ is the dimensionless frequency ($T_{CMB}$ is the temperature of the
present-day cosmic microwave background --- 2.728 K; Fixsen et al. 1996, and $\nu$ is the
observing frequency).  We ignore the complication of relativistic effects, which only
modify the Compton parameter of the hottest clusters by a few percent (Itoh et al. 1998;
Nozawa et al. 2000).

We use the best-fit $\beta$ model parameters [$y_0$ ($\Delta T_0$), $\beta$, and
$\theta_c$] from the literature (discussed below) to reconstruct $y(\theta)$.  The surface 
brightnesses are then numerically integrated within the central $1\arcmin$ (the result is 
symbolized by $y_{int}$) via

\begin{equation}
y_{int}(\leq \theta = 1\arcmin) = 2 \pi \int_0^{\theta = 1\arcmin} y(\theta') \theta'
d\theta'
\end{equation}

\noindent and, finally, converted into a frequency-independent flux density through

\begin{equation}
S_{\nu,arc}/f_{\nu} = y_{int} \biggr[\frac{2(k T_{CMB})^3}{(h c)^2} \biggl]
\end{equation}

\noindent where $f_{\nu}$ is a function of the dimensionless frequency $x$ and is
defined in MBHB03.

Ideally, the three $\beta$ model parameters would be determined by fitting the model to the 
SZ effect data.  However, current SZ effect data cannot tightly constrain these parameters 
when all three are left to vary independently (e.g., Carlstrom et al. 1996; Grego et al. 
2000; 2001; Pointecouteau et al. 2001; 2002).  This problem is often circumvented by 
adopting the best-fit values of $\beta$ and $\theta_c$ (the ``shape'' parameters) 
determined from fitting to the {\it X-ray} surface brightness profile of the cluster and 
leaving only the normalization, $y_0$, to be determined from fitting to the SZ effect data 
(e.g., Pointecouteau et al. 1999; 2001; 2002; Patel et al. 2000; Jones et al. 2003; 
LaRoque et al. 2003; Grainge et al. 2002b).  In the case of clusters with moderate 
redshifts (which 
is a good description of most of the clusters in Table 1), X-ray data still provide better 
constraints on $\beta$ and $\theta_c$ than do SZ effect data.  Thus, in the cases of A1914 
and RXJ2228+2037, we use the X-ray-determined values of these parameters to calculate 
$S_{\nu,arc}/f_{\nu}$.  However, a better approach is to use all available data (both the SZ 
effect and X-ray data) to constrain these parameters.  In estimating Hubble's constant from 
a sample of 18 distant clusters, Reese et al. (2002) did just this.  The shape parameters, 
central X-ray surface brightness, and $y_0$ were all determined simultaneously by using a 
joint maximum-likelihood analysis of both SZ effect and X-ray data.  Because their sample 
is large (in fact, it is the largest sample of SZ effect clusters observed to date) and 
homogeneously analysed and their method takes advantage of all ICM imaging data, we 
preferentially use the values of $y_0$, $\beta$, and $\theta_c$ measured by Reese et al. 
(2002) when multiple measurements for a particular cluster are available (which is the case 
for roughly half of the clusters listed in Table 1).  We note that the agreement 
between different studies is reasonably good (estimates of $y_0$ differ by $\lesssim$ 20\% 
from study to study, e.g., Holzapfel et al. 1997; Jones et al. 2003).  Using the best-fit 
parameters of Reese et al. (2002), we calculate $S_{\nu,arc}/f_{\nu}$ for these 18 clusters.  
Published values for the shape parameters of the two remaining clusters in Table 1, A2204 
and Zwicky 3146, are not available.

To calculate the uncertainty associated with the SZ effect flux density, we vary $y_0$, 
$\beta$, and $\theta_c$ within their allowable ranges (each of the three free parameters has 
an associated statistical uncertainty) to determine the maximum and minimum possible flux 
density of the cluster.  This method actually {\it overestimates} the uncertainty associated 
with the flux density as there is a known correlation between the $\beta$ and 
$\theta_c$ parameters for current SZ effect data (e.g., Grego et al. 2000; 2001).

We are also interested in mapping out correlations between SZ effect and X-ray observables.  
In Table 2, we list the total dark matter masses within $r_{500}$ [$M(r_{500})$], the mean 
emission-weighted gas temperatures ($T_X$), and bolometric X-ray luminosities ($L_X$) of 
these clusters.  We also list whether the cluster has a sharp centrally-peaked X-ray 
surface brightness profile which presumably indicates of the presence of a cooling flow (CF 
indicates the cluster is a ``cooling flow cluster'', NCF indicates that it is not; 
Allen 2000; Reese et al. 2002) and the references for $M(r_{500})$, $T_X$, and $L_X$ 
(respectively).  Below, we discuss the X-ray data in columns 2-4 of Table 2.

For the total dark matter masses within $r_{500}$ (col. 2), we turn to the study of Ettori 
\& Fabian (1999).  By constructing X-ray surface brightness profiles that are based on the 
``universal'' dark matter density (NFW)  profile (Navarro et al. 1997) and comparing these 
to data from {\it ROSAT}, these authors were able to model the underlying dark matter 
density for nine of the clusters listed in our Table 1.  We use their best-fit parameters 
(listed in their Tables 1 and 2) to recover the dark matter density profiles of these 
clusters.  We take care to properly scale these parameters for our assumed cosmology.  We 
then integrate the density profiles out to $r_{500}$ to determine the total dark matter 
mass within that radius.  We are unable to estimate the uncertainty on the masses, 
as there is no reported uncertainty for the best-fit NFW parameters of these clusters.

The most recently determined values for the ICM temperatures of these clusters are 
also listed in Table 2 (col. 3).  Of the 21 clusters with reported temperatures, 16 have 
temperatures determined via fitting to {\it ASCA} X-ray spectral data (Allen 2000; White 
2000; Novicki, Sornig, \& Henry 2002; Jones et al. 2003) and five have measurements based 
on fits to new {\it Chandra} X-ray spectral data (Markevitch \& Vikhlinin 2001; Machacek et 
al. 2002; Vikhlinin et al. 2002).  The cluster RXJ2228+2037 does not have a temperature 
deduced from X-ray spectral analysis, although its central temperature was 
estimated by a combined analysis of X-ray and SZ effect {\it imaging} data (Pointecouteau 
et al. 2002).  In the interest of homogeneity, however, we do not include this cluster in 
our analysis of scaling relations involving the emission-weighted gas temperatures because 
it is unclear how the reported temperature is related to the X-ray emission-weighted 
temperature.  For the clusters that have sharp centrally-peaked X-ray surface 
brightnesses and apparently harbor massive cooling flows (Allen 2000; Reese et al. 
2002; see col. 5), we use cooling flow-corrected temperatures.  The temperatures were 
corrected by fitting the X-ray spectra with a multi-phase plasma model that explicitly takes 
into account the cooler emission from the cluster core (Allen 2000).  We note that for the 
clusters in this sample, cooling flow correction only slightly increases the temperature of 
a cluster (by about 10\% or 1 keV) and does not significantly affect our results.

Finally, column 4 lists the bolometric X-ray luminosities of these clusters.  We 
preferentially select data that has published uncertainties (e.g., White, Jones, \& Forman 
1997; Novicki, Sornig, \& Henry 2002).  For data without published uncertainties, we select 
on the basis of X-ray satellite in the following order: {\it Chandra/XMM-Newton}, {\it 
ASCA}, and {\it ROSAT/Einstein}.  Of the 17 clusters that have published 
bolometric luminosities, seven were determined using {\it Einstein} imaging data (White, 
Jones, \& Forman 1997), one was determined using {\it ROSAT} imaging data (Allen \& Fabian 
1998), seven were determined using {\it ASCA} imaging data (Mushotzky \& Scharf 1997; 
Novicki, Sornig, \& Henry 2002), and two were determined using new {\it Chandra} imaging 
data (Vikhlinin et al. 2002).  Unfortunately, cooling flow-corrected luminosities (which 
normally entails excising the central 100-200 kpc of the CF cluster; e.g., Markevitch 1998; 
Vikhlinin et al. 2002) were not available for the CF clusters in Table 2.

Tables 1 and 2 comprise the largest compiled sample of high redshift SZ effect clusters to 
date.  We note that the current compilation is not statistically complete or homogeneous, as
discussed in Reese et al (2002).  Clusters with bright radio point sources were avoided, for 
the most part, and targets were selected primarily on the basis of X-ray luminosity.  This 
compilation can not be simply defined as being flux-limited or luminosity-limited, for 
example, and it is not complete over the range of redshifts studied.  Without a simple 
selection function it is difficult to assess possible systematic effects that could
be introduced by not having a sample that adequately reflects the demographics of the 
general population of galaxy clusters.

Below, we compare the data in Tables 1 and 2 to our theoretically predicted scaling 
relations (see Table 1 of MBHB03) and attempt to determine if the data require an entropy 
floor (similar to that of X-ray data of nearby clusters) and, if so, what is the level of 
that entropy floor.

\section{Results}

\subsection{Comparing Theory to Observations}

A detailed analysis of how the presence of excess entropy alters SZ effect scaling 
relations was presented in the companion paper MBHB03.  These relations were derived 
using the entropy injection (preheated) models developed by Babul et al. (2002).  
Following standard practice, we fitted simple power-law models to the theoretical scaling 
relations.  Since a number of the relations were quite sensitive to redshift and existing 
SZ effect data span a wide range in $z$ (see Table 1), we fitted power-law models which 
were an explicit function of both redshift and entropy floor level ($K_0$) to the 
theoretical relations; i.e., for arbitrary cluster parameters $X$ and $Y$ (e.g., $y_0$ 
and $T_X$)

\begin{equation}
\log_{10}{Y} = a(K_0,z) \log_{10}{X} + b(K_0,z)
\end{equation}

With these relations, it is possible to quickly and accurately compute how various cluster 
properties scale with the SZ effect of the Babul et al. (2002) model clusters at any 
redshift between $0.1 \lesssim z \lesssim 1$ and with any entropy floor in the range $100$ 
keV cm$^2$ $\lesssim K_0 \lesssim 700$ keV cm$^2$ (see the Table 1 of MBHB03).  This is 
the only way a fair comparison can be made between the existing data and the theoretical 
models.

We compare the theoretical scaling relations to the observational data compiled in Tables 
1 and 2 with a $\chi^2$ statistic

\begin{equation}
\chi^2 = \sum_{i=1}^{N} \frac{[\log{Y_i} - a(K_0,z) \log{X_i} - 
b(K_0,z)]^2}{\sigma^2_{Y_{i}} + a(K_0,z)^2 \sigma_{X_{i}}^2} 
\end{equation}

\noindent where $\sigma^2_{Y_{i}}$ and $\sigma^2_{X_{i}}$ are the measurement errors in  
log-space and are estimated by multiplying the measurement errors in linear-space by the 
weighting factors $(Y_{i} \log{e})^{-1}$ and $(X_{i} \log{e})^{-1}$, respectively (see, 
e.g., Press et al. 1992). 

We specify the optically-determined redshifts for each of the clusters and then determine 
the best-fit value of entropy floor level, $K_0$, by minimizing the $\chi^2$ (assuming that 
the value of $K_0$ is the same for all clusters).  Unless stated otherwise, quoted error 
bars are for the 95.4\% ($2\sigma$) confidence level (corresponding to $\Delta \chi^2 = 4$).  
We assess the quality of the fit by calculating the reduced-$\chi^2$ ($\chi^2_{\nu}$).  For 
the self-similar model (whose scaling relations depend on $z$ only; i.e., there are no free 
parameters), we calculate the $\chi^2$ and $\chi^2_{\nu}$ and compare this with that of the 
best-fit entropy floor model.

Below, we examine seven different scaling relations involving the two primary SZ 
effect observables (the central Compton parameter and the integrated flux density).
First, we explore the trend that exists between the two SZ effect observables.  This 
is an extremely interesting test since it can potentially be measured entirely 
independent of X-ray observations.  In fact, in \S 4, we show that future data from the 
{\it SZA} and the upgraded {\it OVRO} array will allow one to constrain the central 
entropy distribution of clusters all the way out to $z \sim 2$ using this relation.
The next two trends that we study, between the two SZ effect observables and the mass of 
a cluster, are also interesting because they too can potentially be measured 
independent of X-ray observations.  Aside from X-ray observations, one can estimate the 
mass of a cluster via gravitational lensing or through galaxy velocity dispersion 
measurements.  We expect that the $y_0-M(r_{500})$ relation, in particular, will be very 
sensitive to the presence of excess entropy.  Finally, the four remaining correlations 
that we look at are between the two SZ effect observables and the two primary X-ray 
observables, i.e., the mean emission-weighted temperature ($T_X$) and the bolometric 
luminosity ($L_X$).  The correlations between the SZ effect observables and the cluster 
temperature are expected to be excellent probes of the central entropy of clusters and, 
as such, we discuss these ahead of those involving $L_X$.  

\subsection{A SZ Effect Only Relation}

We start by examining the correlation between the two SZ effect quantities, 
$S_{\nu,arc}/f_{\nu}$ and $y_0$ (see Fig. 2 of MBHB03).  Even though theoretical 
arguments suggest that this relation is not as sensitive to the presence of excess entropy 
as some of the other relations we will discuss later, it is by far the most interesting.  
As we already mentioned, this relation can potentially be measured entirely through SZ 
effect observations, offering a completely X-ray-independent probe of the intracluster 
gas.  This is discussed in more detail below.

Fitting all 20 clusters in Table 1 for which we have estimates of both 
$S_{\nu,arc}/f_{\nu}$ and $y_0$, we find a best-fit entropy floor level of $K_0 = 
540^{+170}_{-165}$ keV cm$^2$ with a $\chi^2_{\nu}$ = 37.77/19 = 1.99.  A residual plot 
demonstrating the quality of our best-fit entropy floor model is presented in Figure 1 (top 
{\epsscale{1.0}
\plotone{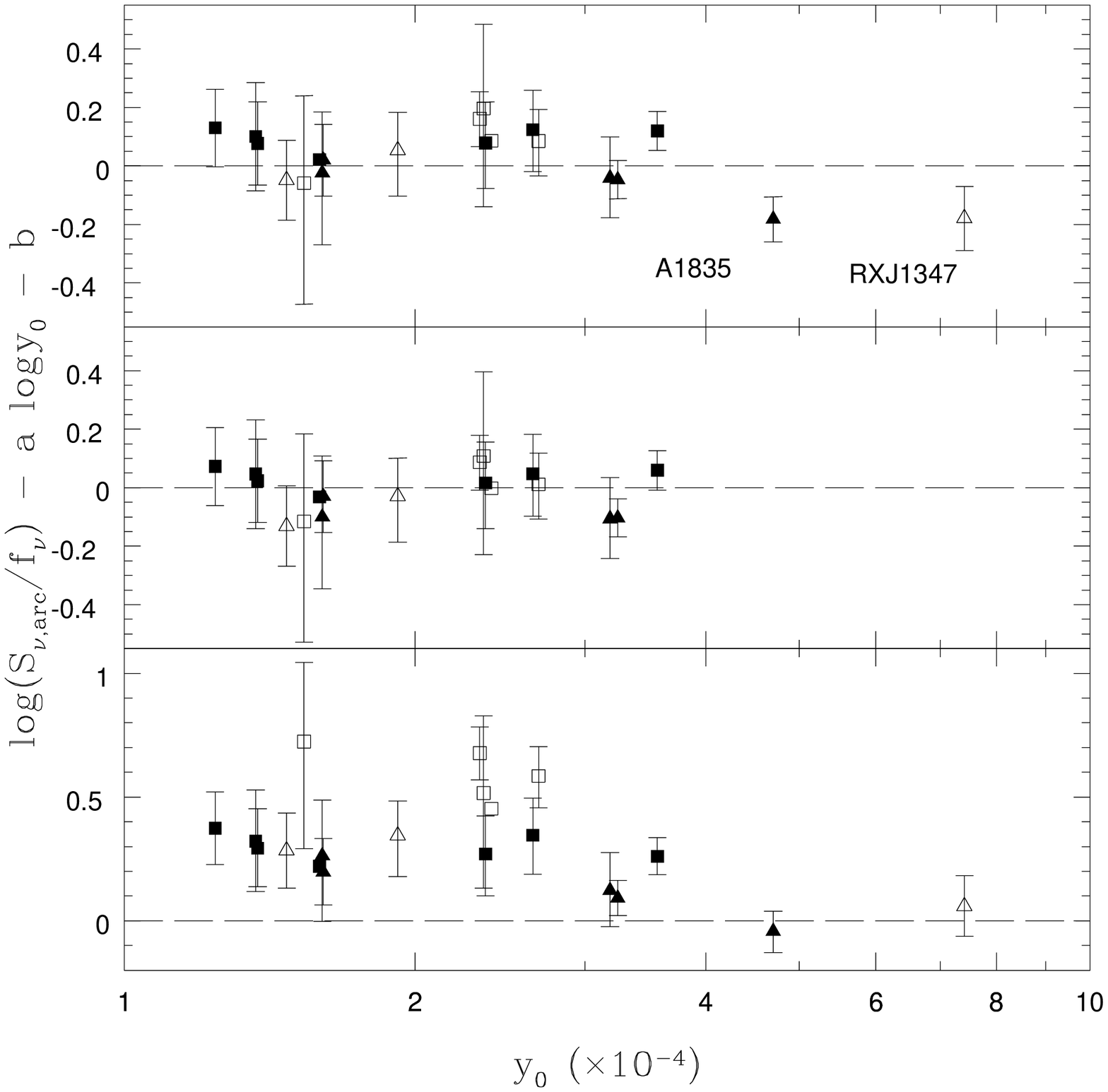}
{Fig. 1. \footnotesize
Residual plots for the $S_{\nu,arc}/f_{\nu}-y_0$ relation.  {\it Top:}
Residuals from a comparison between the entropy floor model with $K_0 = 540$ keV cm$^2$ and
the data in Table 1 ($\chi^2_{\nu}$ = 1.99).  {\it Middle:} Neglecting A1835 and
RXJ1347.5-1145 - for reasons discussed in the text - greatly improves the fit: $K_0 = 570$
keV cm$^2$ with $\chi^2_{\nu} = 0.99$.  {\it Bottom:} Residuals from a comparison
between the self-similar model and the data in Table 1 ($\chi^2_{\nu}$ = 12.21).  Filled
symbols represent low redshift ($z < 0.3$) clusters, open symbols represent high redshift
($z > 0.3$) clusters, triangles represent cooling flow clusters, and squares represent non
cooling flow clusters.  For clarity, error bars for the abscissa are not displayed.
}}
\vskip0.1in
\noindent
panel).  The residuals from a comparison between the self-similar model and the data in 
Table 1 is presented in the bottom panel ($\chi^2_{\nu}$ = 244.15/20 = 12.21).  Note that 
the residuals in the top panel are generally consistent with the zero line (indicating a 
relatively good fit), whereas the residuals in the bottom panel are systematically too 
high, indicating that the self-similar model predicts values of  $S_{\nu,arc}/f_{\nu}$ (for 
a fixed $y_0$) that are systematically lower than observed.  This is equivalent to saying 
that the self-similar model predicts clusters that have SZ effect surface brightness 
profiles that are too centrally-peaked.  The entropy floor models are able to match the 
observations much better because the addition of an entropy floor flattens the pressure 
profiles of clusters (e.g., Fig. 1. of MBHB03) and since the SZ effect is proportional 
to integrated line-of-sight pressure of the cluster, this leads to flatter surface 
brightness profiles.
  
There is little doubt that the best-fit entropy floor model provides a much better 
fit than the self-similar model to the observational data, but the best-fit entropy floor 
model itself does not provide a statistically acceptable fit to the data (note the 
high value of the $\chi^2_{\nu}$).  However, the residuals in the top panel of Figure 
1 clearly show two obvious outliers, A1835 and RXJ1347.5-1145.  Neglecting these two 
clusters, the quality of the fit to the whole sample is significantly improved: we find 
$K_0 = 575^{+150}_{-155}$ keV cm$^2$ with $\chi^2_{\nu} = 16.79/17 = 0.99$.  The residuals 
for this fit are plotted in the middle panel of Figure 1.  This result is  
consistent with that determined from studies of X-ray scaling relations of nearby massive 
clusters, which require $K_0 \gtrsim 300$ keV cm$^2$ (e.g., Tozzi \& Norman 2001; Babul et 
al. 2002; McCarthy et al. 2002).

It is interesting that the two outliers, A1835 and RXJ1347.5-1145, have extraordinarily 
large cooling flow mass deposition rates.  Both apparently deposit several thousand solar 
masses of gas each year and are among the most massive cooling flow clusters known (Allen 
2000).  In fact, a closer inspection of the residuals (particularly in the left-hand panel) 
reveals that there is a small systematic difference between cooling flow clusters 
(triangles) and non cooling flow clusters (squares), even if one neglects these two 
outliers.  While in principle the SZ effect should be less susceptible than the 
X-ray emission to the effects of cooling flows (because SZ effect is proportional to $n_e$, 
whereas the X-ray emission is proportional to $n_e^2$), recall that both of the SZ effect 
quantities used in Figure 1 were inferred through fitting to both X-ray and SZ effect 
data.  Thus, the strong central surface brightness peaks present in the X-ray images of 
these cooling flow clusters will have an impact on the implied SZ effect observables.  
Excluding A1835 and RXJ1347.5-1145, however, there is no statistical evidence for a 
difference in $K_0$ when fitting to the cooling flow and non cooling flow clusters 
separately or to the whole sample.  Therefore, it appears that only the most extreme 
cooling flow clusters could have significantly different entropy histories.  

Because the sample in Table 1 is reasonably large and spans a wide range of redshifts, it 
is possible to use the catalog to get some idea of how the entropy floor level of clusters 
evolved with cosmic time.  We split the sample up into two large redshift bins: (1) ``low'' 
redshift clusters ($z < 0.3$), and (2) ``high'' redshift clusters ($z > 0.3$).  Fitting 
only the low redshift clusters, we derive $K_0 = 505^{+220}_{-210}$ keV cm$^2$ 
($\chi^2_{\nu} = 10.46/10 = 1.05$), which is consistent with the value derived from fitting 
to the entire sample.  The high redshift clusters, however, prefer a slightly elevated 
entropy floor with $K_0 = 640^{+205}_{-215}$ keV cm$^2$ ($\chi^2_{\nu} = 5.84/6 = 0.97$), 
although the difference between the low and high redshift clusters is not statistically 
significant.  Thus, there is no good evidence that $K_0$ evolves significantly out to $z 
\sim 0.7$, at least on the basis of this test.  A stronger test of this hypothesis will 
soon be possible, as the list of high redshift clusters observed through the SZ effect 
is growing rapidly (Carlstrom and Joy and collaborators, for example, have now made 
detections in 21 clusters with $z > 0.45$; Reese et al. 2002).

Finally, to what extent the present $S_{\nu,arc}/f_{\nu}-y_0$ relation studied here is 
independent of previous X-ray results is debatable.  Both $S_{\nu,arc}/f_{\nu}$ and $y_0$ 
were {\it inferred} by fitting a model to the SZ effect surface brightness profiles of 
clusters and, furthermore, this surface brightness model (the isothermal $\beta$ model) had 
two of its three free parameters constrained to be the same as that deduced from X-ray 
observations (Jones et al. 2003; Pointecouteau et al. 2002) or from X-ray and SZ effect 
observations (Reese et al. 2002).  Yet, it is also clear that this test is different from 
any other scaling relation examined to date.  Ideally, both $S_{\nu,arc}/f_{\nu}$ and $y_0$ 
could be measured directly or, failing that, determined from fitting a model to only the 
SZ effect data.  Current SZ effect data alone, however, cannot 
{\epsscale{1.0}
\plotone{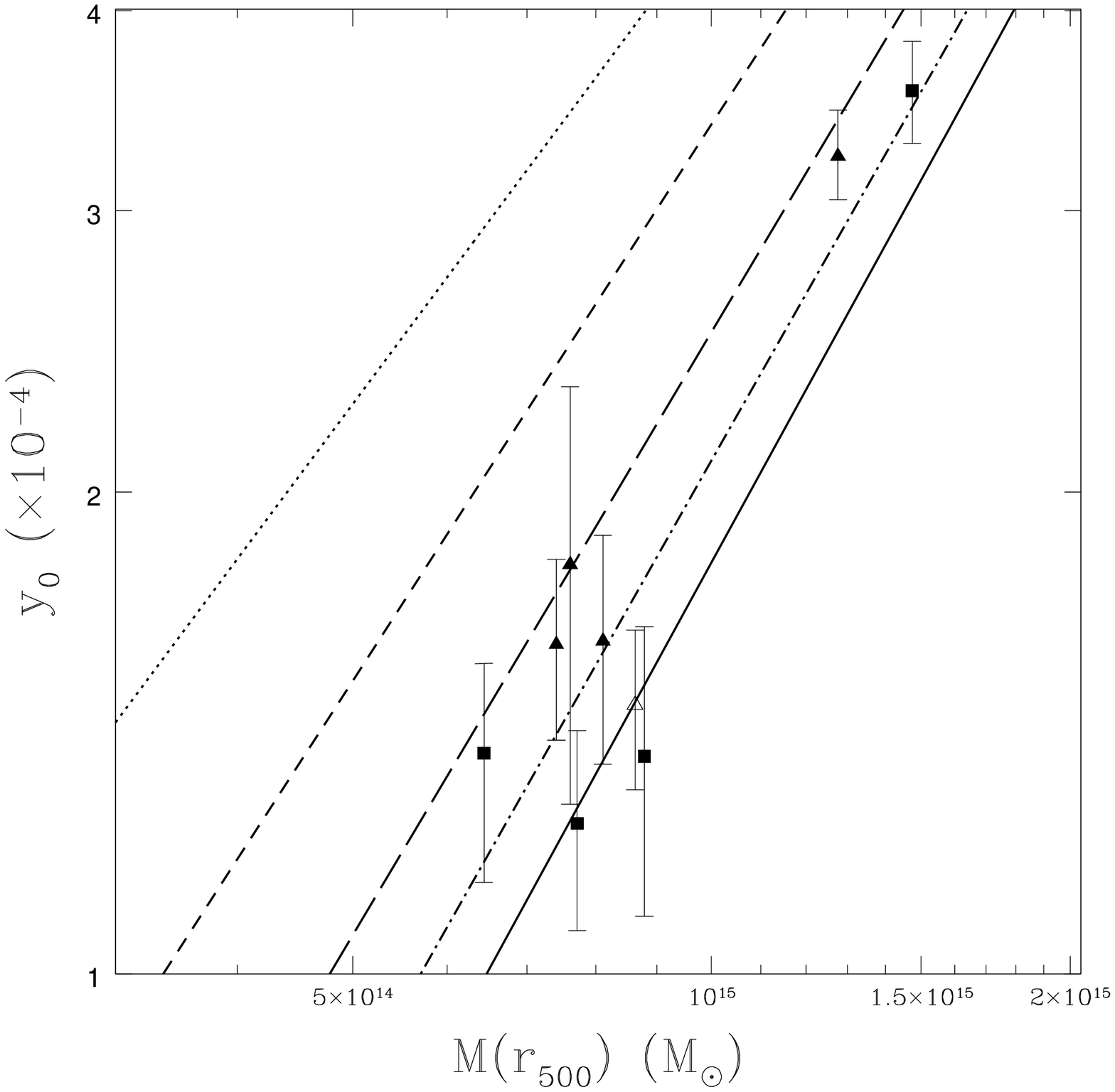}
{Fig. 2. \footnotesize
The observed and predicted $y_0-M(r_{500})$ relations.  The symbols have the
same meaning as in Figure 1.  The dotted, short-dashed, long-dashed, dot-dashed, and solid
lines represent the self-similar and $K_0 =$ 100, 300, 500, and 700 keV cm$^2$ entropy 
floor models, respectively.
}}
\vskip0.1in
\noindent
tightly constrain these 
quantities.  In \S 4, we show that the high quality data to be produced by the upcoming 
{\it SZA} and the (soon to be) upgraded {\it OVRO} array will make it possible to accurately 
measure these quantities for massive clusters at virtually any redshift and without having 
to use any X-ray results. 

\subsection{The SZ effect-$M(r_{500})$ Relations}  

The next set of scaling relations we examine are the SZ effect - cluster mass 
relations.  Theoretical arguments suggest that these relations should be very sensitive to 
the presence of entropy floor, at least when the SZ effect is measured near the cluster 
center (e.g., $y_0$).  Interestingly, these trends too can also potentially be measured 
independent of X-ray results.  Future SZ effect observations will allow one to estimate 
$y_0$ and $S_{\nu,arc}/f_{\nu}$ accurately purely through SZ effect surface brightness
profiles (see \S 4), while both strong and weak lensing are increasingly being used to 
measure
the mass profiles of clusters out to radii comparable in size to that of $r_{500}$ (e.g.,
Clowe \& Schneider 2001).  At present, however, only a few of the clusters in our sample 
have been weighed using lensing.

In Figure 2, we plot the observed $y_0-M(r_{500})$ relation.  This is superimposed on the 
predicted $z = 0.2$ $y_0-M(r_{500})$ relations for the self-similar model (dotted line) and 
the $K_0$ = 100 (short-dashed), 300 (long-dashed), 500 (dot-dashed), and 700 keV cm$^2$ 
(solid) entropy floor models.  Because the majority of the clusters with published masses 
in Tables 1 and 2 lie in a narrow redshift range around $z \sim 0.2$, it is possible to 
{\it qualitatively} compare the theoretical models to the data ``by eye'' before using the 
quantitative method outlined in \S 3.1.  A simple, neat, and fair qualitative comparison 
was not possible for the previous relation ($S_{\nu,arc}/f_{\nu}-y_0$), as the data spanned 
{\epsscale{1.0}
\plotone{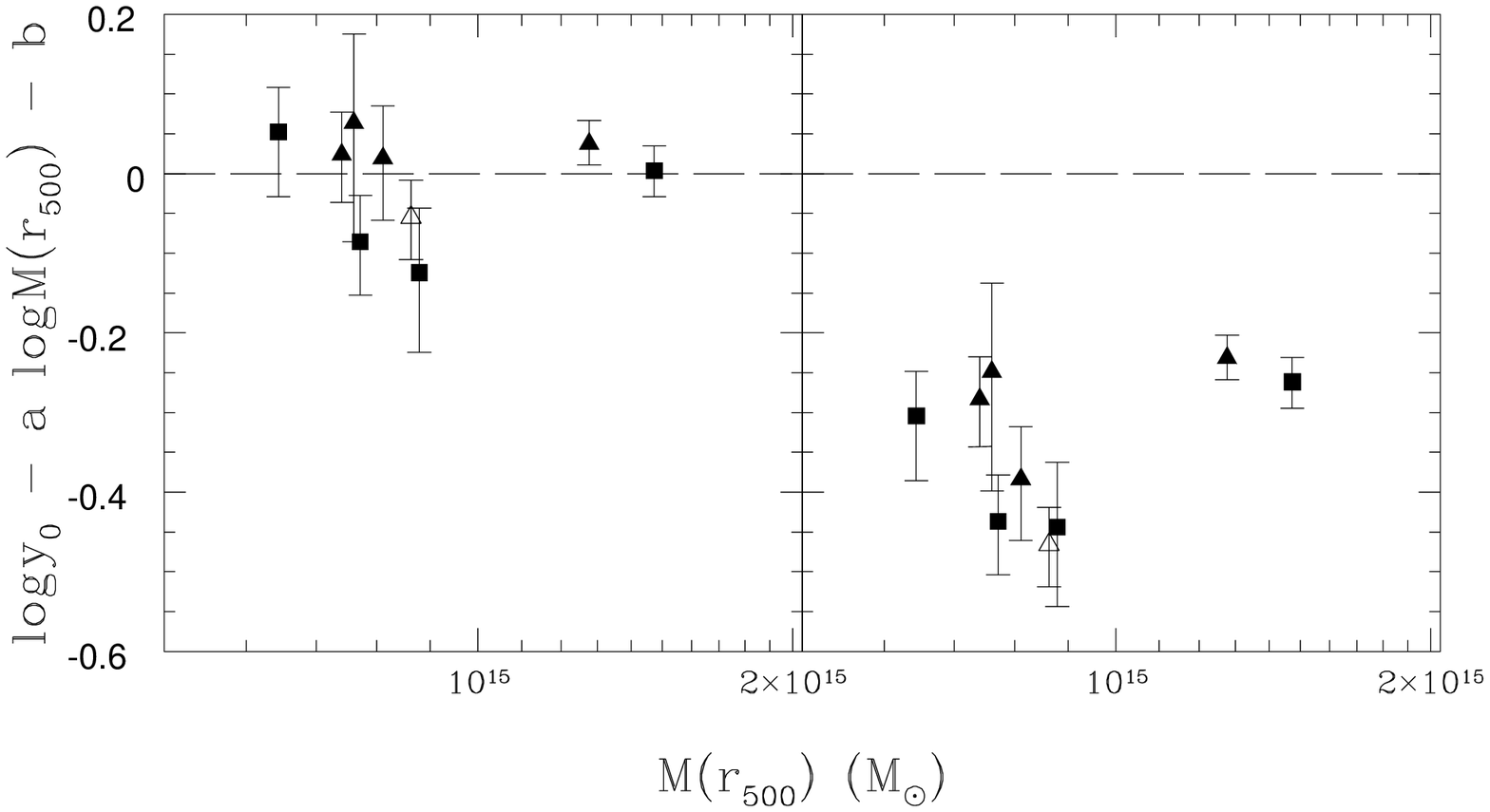}
{Fig. 3. \footnotesize
Residual plots for the $y_0-M(r_{500})$ relation.  {\it Left:} Residuals from a
comparison between the entropy floor model with $K_0 = 500$ keV cm$^2$ and the
observational data ($\chi^2_{\nu}$ = 1.00).  {\it Right:} Residuals from a comparison
between the self-similar model and the observational data ($\chi^2_{\nu}$ = 41.80).  The
symbols have the same meaning as in Figure 1.  For clarity, error bars for the abscissa are
not displayed.
}}
\vskip0.1in
\noindent
a wide range of redshifts and because that relation is especially sensitive to $z$ 
($S_{\nu}$ scales as $1/D_a^{2}$).

By visual inspection of Figure 2, it is obvious that only the high entropy floor models 
($K_0 \gtrsim 300$ keV cm$^2$) provide a reasonable fit to the observational data.  In 
addition, the observed correlation does not seem to depend on the cooling flow status or 
redshift of the clusters, although the sample is far too small to make any robust 
conclusions to this effect.  Fitting all nine clusters with the method outlined in \S 
3.1, our best-fit entropy floor level is $K_0 = 500^{+65}_{-65}$ keV cm$^2$ with 
$\chi^2_{\nu}$ = 8.02/8 = 1.00.  This is consistent with the results derived in \S 3.2 and 
with X-ray observations of nearby massive clusters.  A plot of the residuals between the 
data and the $K_0 = 500$ keV cm$^2$ model is shown in the left-hand panel of Figure 3.  Also 
shown (right-hand panel) are the residuals of a comparison between the data and the 
self-similar model ($\chi^2_{\nu}$ = 376.05/9 = 41.80).  The residuals for the $K_0 = 500$ 
keV cm$^2$ model display a tight scatter about the zero line, while the residuals for the 
self-similar model indicate that $y_0$ is observed to be much lower [for a fixed value of 
$M(r_{500})$] than predicted by this model.  The entropy floor models with $K_0 \sim 500$ 
keV cm$^2$ are able to provide a good match to the data because the addition of an entropy 
floor reduces the gas pressure near the centers of clusters (see MBHB03 for a detailed 
discussion).  This, in turn, reduces the magnitude of $y_0$.  The mass within $r_{500}$, 
however, is unaffected by the modification of the gas entropy.

Although the available $y_0-M(r_{500})$ data exhibit only a very small amount of scatter 
about the $K_0 \sim 500$ keV cm$^2$ relation (and the $\chi^2_{\nu}$ indicates a very 
good fit), the estimated error bars on our best-fit value of $K_0$ from this relation 
are almost certainly too small.  We say this because (1) we were unable to calculate 
any uncertainty for $M(r_{500})$ as there were no published error bars for the best-fit NFW 
parameters for the clusters in Figures 2 and 3, and (2) the sample is too small to get any 
kind of a handle on the systematic errors associated with the cluster masses.  For example, 
estimates of cluster masses from gravitational lensing very often differ from those 
determined from X-ray data (sometimes by up to a factor of two; e.g., Miralda-Escud\'{e} \& 
Babul 1995; Wu \& Fang 1997; 
{\epsscale{1.0}
\plotone{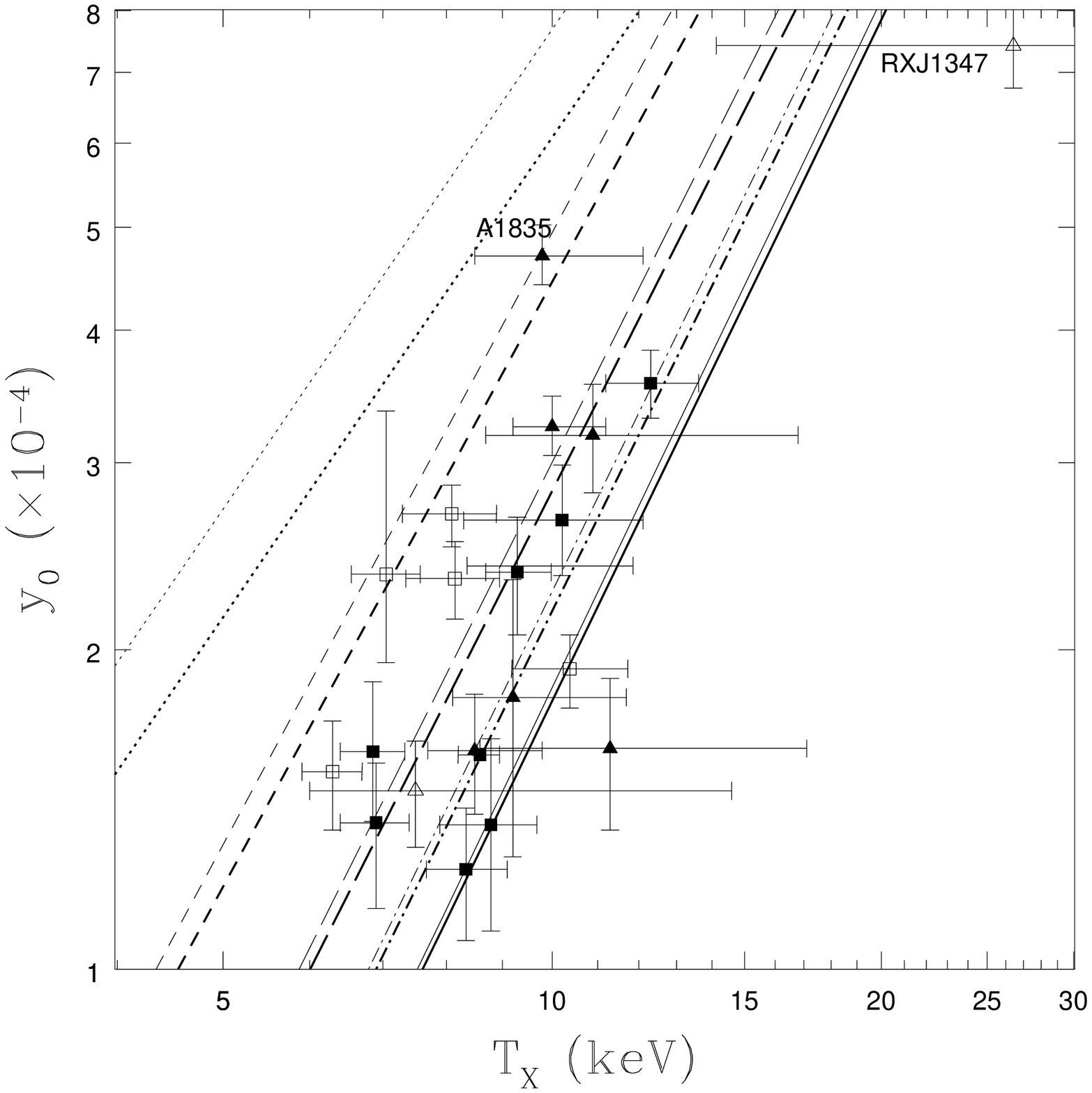}
{Fig. 4. \footnotesize
The observed and predicted $y_0 - T_X$ relations.  The symbols have the same
meaning as in Figure 1.  The dotted, short-dashed, long-dashed, dot-dashed, and solid
lines represent the self-similar and $K_0 =$ 100, 300, 500, and 700 keV cm$^2$ entropy
floor models, respectively.  Thin lines are for $z = 0.2$ and thick lines are for $z =
0.5$.  As discussed in the text, the models with $K_0 \sim 300$ keV cm$^2$ provide the
best description of the observations.
}}
\vskip0.1in
\noindent
Ota et al. 2002) and it is not yet clear why this happens to 
be the case.  There could be a systematic problem with the X-ray determined masses (e.g., 
Allen 1998).

Aside from the $y_0-M(r_{500})$ relation, we also explore the 
$S_{\nu,arc}/f_{\nu}-M(r_{500})$ trend.  This trend could provide a consistency check of the  
results discussed immediately above.  Unfortunately, the available SZ effect flux density - 
mass data do not constrain $K_0$.  This is not completely unexpected since the 
$S_{\nu,arc}/f_{\nu}-M(r_{500})$ relation is much less sensitive than the $y_0-M(r_{500})$ 
relation to the entropy floor level of clusters (MBHB03).  Future data from, for 
example, the {\it SZA} will allow for much more precise determinations of the SZ effect 
flux densities of clusters and we expect that the future $S_{\nu,arc}/f_{\nu}-M(r_{500})$ 
relation will place much tighter constraints on $K_0$.

\subsection{The SZ effect$-T_X$ Relations}

Plotted in Figure 4 is the observed $y_0-T_X$ relation.  In addition to the observational 
data, we have also plotted the $z = 0.2$ (thick lines) and $z = 0.5$ (thin lines) 
theoretical relations.  The various line types (e.g., solid, dotted) have the same 
meanings as in Figure 2.  Again, because the theoretical relations are not a strong 
function of redshift, a preliminary visual comparison of the data and models is feasible.

First, it is clear that the data indicate that there is a fairly tight correlation between
a cluster's central Compton parameter and its emission-weighted gas temperature.  This
correlation seems to hold true irrespective of whether the clusters have cooling flows or
not.  This may be expected since we have used cooling flow-corrected temperatures.
However, even without this correction the data still 
{\epsscale{1.0}
\plotone{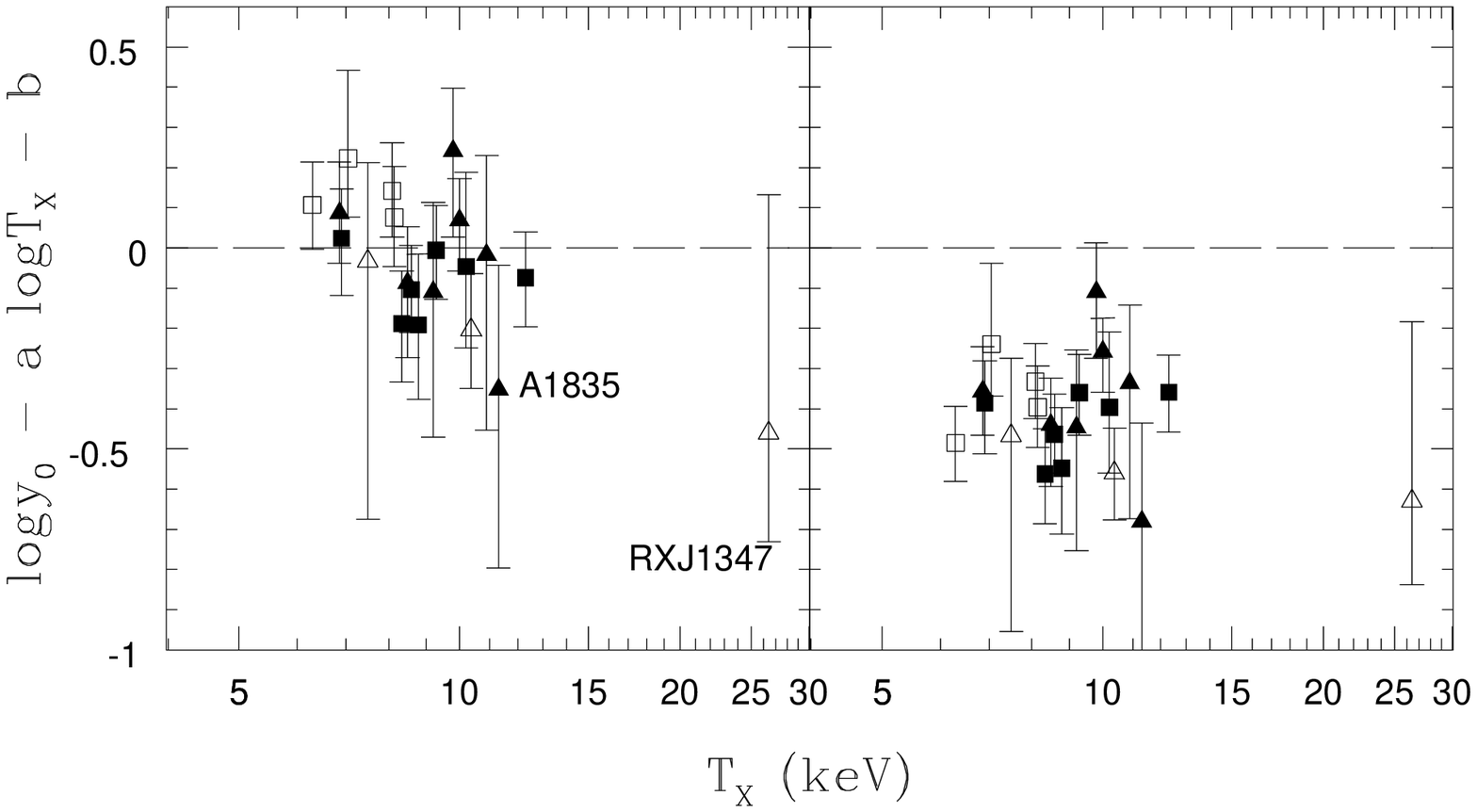}
{Fig. 5. \footnotesize
Residual plots for the $y_0 - T_X$ relation.  {\it Left:} Residuals from a
comparison between the entropy floor model with $K_0 = 300$ keV cm$^2$ and the
observational data ($\chi^2_{\nu}$ = 1.32).  Neglecting A1835 and RXJ1347.5-1145, we find
a best fit of $K_0 = 305$ keV cm$^2$ with $\chi^2_{\nu}$ = 1.24.  {\it Right:} Residuals
from a comparison between the self-similar model and the observational data
($\chi^2_{\nu}$ = 21.41).  The symbols have the same meaning as in Figure 1.  For clarity,
error bars for the abscissa are not displayed.
}}
\vskip0.1in
\noindent
exhibit a tight trend.  For the
clusters in this sample, cooling flow correction only slightly increases the temperature 
of a cluster (by about 10\% or 1 keV) and, therefore, does not significantly affect the
results.  There is also a hint of a slight systematic difference in the $y_0-T_X$ 
relations for low and high redshift clusters (filled and open symbols, respectively).  In 
particular, the relation for the high redshift clusters has a normalization that is 
slightly higher than that of the relation for low redshift clusters (ignoring RXJ1347, 
whose temperature is quite uncertain).  This is discussed in more detail below.

Comparing the various theoretical relations to the observational data, it is apparent that
the standard self-similar model predicts a $y_0-T_X$ relation which is a poor match to the
data. In particular, the normalizations of the self-similar relations are roughly 2.5 
times larger than observed (over the range 6 keV $\lesssim T_X \lesssim 12$ keV).  The 
entropy floor models with $K_0 \sim 300$ keV cm$^2$, on the other hand, provide a very 
good qualitative fit to the data.  In MBHB03, we found that injecting the ICM with
non-gravitational entropy tends to decrease $y_0$ and, at the same time, increase $T_X$.
This reduces the normalization of the predicted trend between these two quantities and, as
is apparent from Figure 4, brings close agreement between theory and observations.

Fitting all 21 clusters in Tables 1 and 2 via the method outlined in \S 3.1, we obtain a
best-fit entropy floor of $K_0 = 300^{+80}_{-60}$ keV cm$^2$ with $\chi^2_{\nu} = 26.36/20
= 1.32$.  The residuals of a comparison between the $K_0 = 300$ keV cm$^2$ entropy floor
model and the self-similar model with the observation data are plotted in Figure 5.  Note
the striking separation between the residuals of the two models.  Theoretically, the
$y_0-T_X$ relation is the most sensitive to the entropy floor level of any and all scaling
relations that we have explored.

If we exclude A1835 and RXJ1347.5-1145 from the fit, the entropy floor is essentially
unchanged ($K_0 = 305^{+80}_{-65}$ keV cm$^2$) but the fit is improved ($\chi^2_{\nu} =
22.37/18 = 1.24$).  In addition, we verify that splitting the sample by cooling flow
status does not significantly modify the best-fit value 
{\epsscale{1.0}
\plotone{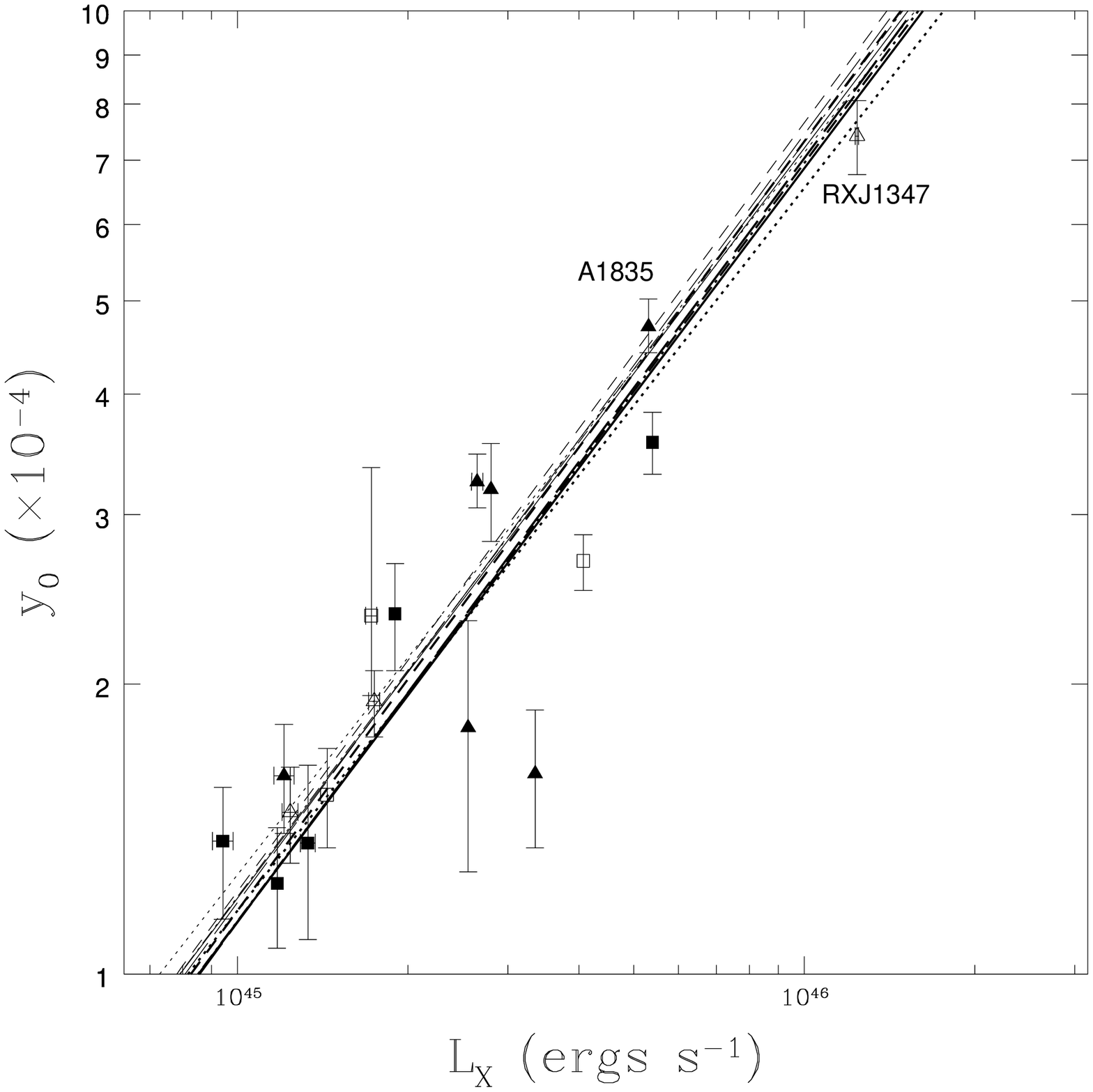}
{Fig. 6. \footnotesize
The observed and predicted $y_0-L_X$ relations.  The symbols have the same
meaning as in Figure 1.  The dotted, short-dashed, long-dashed, dot-dashed, and solid lines
represent the self-similar and $K_0 =$ 100, 300, 500, and 700 keV cm$^2$ entropy floor
models, respectively.  Thin lines are for $z = 0.2$ and thick lines are for $z = 0.5$.
}}
\vskip0.1in
\noindent
of $K_0$.  As mentioned above,
there appears to be a slight difference in the normalizations of the $y_0-T_X$ relations 
for low and high redshift clusters.  Is this difference significant?  If we restrict the 
fit to low redshift ($z < 0.3$) clusters only, we find a best-fit value of $K_0 = 
370^{+130}_{-95}$ keV cm$^2$ ($\chi^2_{\nu} = 9.69/12 = 0.81$).  This is consistent with 
the best-fit for the whole sample.  Fitting only high redshift ($z > 0.3$) clusters we 
find a best-fit value of $K_0 = 220^{+100}_{-70}$ keV cm$^2$ ($\chi^2_{\nu} = 7.64/5 = 
1.53$).  Therefore, there is only a marginal statistical difference between the best-fit 
values of $K_0$ from the low and high redshift clusters.  A very similar result is deduced 
from an examination of the $S_{\nu,arc}/f_{\nu}-T_X$ relation (which is not as sensitive 
as the $y_0 - T_X$ relation to the entropy floor; MBHB03), where the low redshift clusters 
are best fit by entropy floor models with $K_0 \sim 300$ keV cm$^2$ while the high 
redshift clusters are best fit by entropy floor models with $K_0 \sim 200$ keV cm$^2$.

\subsection{The SZ effect$-L_X$ Relations}

The theoretical $y_0-L_X$ relations have the interesting properties that they do not
evolve strongly with redshift and, more importantly, are virtually insensitive to the
presence of an entropy floor.  The reason why this happens to be the case is not because the
individual quantities, $y_0$ and $L_X$, are unaffected by an entropy floor (on the contrary,
they are greatly modified), but because they are both affected in a very similar manner. 
Therefore, this correlation is less than ideal when it comes to probing the 
non-gravitational entropy of galaxy clusters.  However, the relation is still 
of great interest because it provides a valuable consistency check of the other scaling 
relations studied here.  

Figure 6 is a plot of the observed $y_0-L_X$ relation.  This 
{\epsscale{1.0}
\plotone{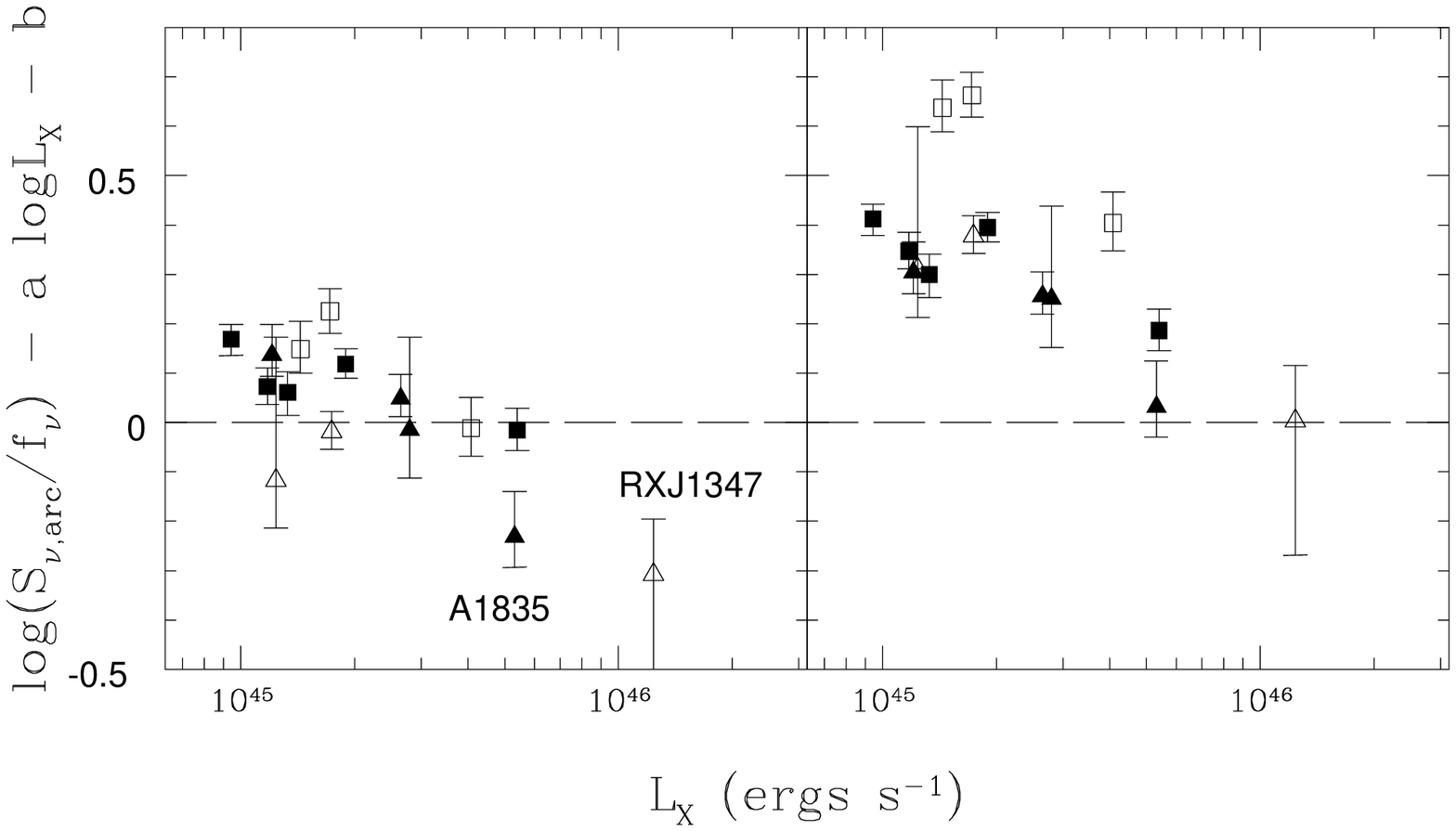}
{Fig. 7. \footnotesize
Residual plots for the $S_{\nu,arc}/f_{\nu}-L_X$ relation.  {\it Left:}
Residuals from a comparison between the entropy floor model with $K_0 = 310$ keV cm$^2$ and
the observational data ($\chi^2_{\nu}$ = 4.27).  Neglecting the two massive ``cooling 
flow''
clusters, A1835 and RXJ1347.5-1145, we converge on $K_0 = 385$ keV cm$^2$ as the best-fit
model with $\chi^2_{\nu}$ = 1.15.  {\it Right:} Residuals from a comparison between the
self-similar model and the observational data ($\chi^2_{\nu}$ = 17.30).  The symbols have
the same meaning as in Figure 1.  For clarity, error bars for the abscissa are not
displayed.
}}
\vskip0.1in
\noindent
is superimposed on the 
predicted $z = 0.2$ (thick lines) and $z = 0.5$ (thin lines) relations for the self-similar 
model (dotted) and the $K_0$ = 100 (short-dashed), 300 (long-dashed), 500 (dot-dashed), and 
700 keV cm$^2$ (solid) entropy floor models.  Since the theoretical $y_0-L_X$ relations do 
not evolve strongly with redshift (compare the thin and thick lines), a qualitative ``by 
eye'' comparison is possible.

A clear correlation between a cluster's central Compton parameter and its bolometric X-ray 
luminosity is apparent in Figure 6.  The relation is quite tight and follows the 
predicted trends.  Even the two massive cooling flow clusters A1835 and RXJ1347.5-1145 seem 
to follow the predicted trends.  Thus, the observed $y_0-L_X$ relation gives us confidence 
that our general understanding of ICM is basically correct.  That being said, the quality 
of the fit is not great in a statistical sense ($\chi^2_{\nu} = 67.36/16 = 4.21$).  The 
high value of $\chi^2_{\nu}$ undoubtedly arises from the random scatter present in the 
observed relation.  The presence of this scatter may be inconsistent with theoretical 
predictions.  We note, however, that observational systematic uncertainties for $y_0$ 
and $L_X$ are on the order of 10\% each (Reese et al. 2002).  In addition, Mushotzky \& 
Scharf (1997) have found that estimates of the X-ray luminosities of clusters can vary from 
study to study by up to 20\%, most likely attributable to different measurement 
techniques.  Conservatively estimating the {\it total} uncertainty on $L_X$ to be 20\%, we 
find a significantly improved fit of $\chi^2_{\nu} = 19.76/16 = 1.24$.  As expected, the 
entropy floor level is not constrained by the data.    

While the $y_0-L_X$ relation is insensitive to the entropy floor level, the  
$S_{\nu,arc}/f_{\nu}-L_X$ relation is not.  Fitting all 15 clusters for which we have 
estimates of both the luminosity and SZ effect flux density, we find a best-fit entropy 
floor level of $K_0 = 310^{+70}_{-70}$ keV cm$^2$.  However, the fit is not a good one, as 
is evident from the residuals plotted in Figure 7 and the calculated reduced-$\chi^2$ 
($\chi^2_{\nu} = 34.34/14 = 4.27$).  The two massive cooling flow clusters A1835 and 
RXJ1347.5-1145 are obvious outliers.  Ignoring these two clusters, we obtain $K_0 = 
385^{+75}_{-70}$ keV cm$^2$ ($2\sigma$ error bars) and a significantly improved fit 
($\chi^2_{\nu} = 13.79/12 = 1.15$).  This best-fit value of $K_0$ is consistent with the 
results of \S 3.2 and \S 3.3 and also with X-ray observations of nearby massive clusters.  
Splitting the sample into two redshift bins ($< 0.3$ and $> 0.3$), we also find there to be 
no difference in the entropy floors of ``nearby'' and ``distant'' galaxy clusters.  This is 
the same as was found for the $S_{\nu,arc}/f_{\nu}-y_0$ relation.
	
\subsection{Summary of Scaling Relations}

Every single SZ effect scaling relation that we have examined is consistent with or 
requires a high value for the entropy floor level, $K_0$.  In fact, several of the 
trends, such as the $y_0-T_X$, $y_0-M(r_{500})$, and $S_{\nu,arc}/f_{\nu}-y_0$ relations, 
rule out the standard self-similar model at many sigma.  Neither of the relations show 
any convincing evidence for strong evolution in $K_0$ out to the limit to which our 
sample extends ($z \sim 0.7$).  

It is interesting that the estimates of $K_0$ from the various relations do not always 
agree.  For example, the best-fit entropy floors from the $y_0-T_X$ and 
$S_{\nu,arc}/f_{\nu}-T_X$ trends are consistent with results from studies of X-ray 
scaling relations of nearby massive clusters (e.g., Babul et al. 2002) and the results of 
our SZ effect-luminosity relations but are marginally lower than the results from our 
$S_{\nu,arc}/f_{\nu}-y_0$ and $y_0-M(r_{500})$ relations.  A conservative estimate of the 
{\it true} value of $K_0$, however, must fall in between the results of each of the 
individual relations; i.e., $300$ keV cm$^2$ $\lesssim K_0 \lesssim 600$ keV cm$^2$.  
This is clearly illustrated in Figure 8 through a plot of $\Delta \chi^2$ vs. $K_0$ for 
the relations we have examined ($\Delta \chi^2 \equiv \chi^2 - \chi^2_{BF}$ where 
$\chi^2_{BF}$ is the $\chi^2$ for the best-fit power-law model for each of the scaling 
relations).  A naive simultaneous fit to all of the relations, ignoring correlated 
variables and errors, yields $K_0 = 430^{+60}_{-55}$ keV cm$^2$ (99\% level), which is 
remarkably similar to that derived solely from X-ray data (c.f. Babul et al. 2002).

At present, it is unclear why some of the scaling relations do not converge on the same 
value of $K_0$.  A more detailed analysis of this would require taking into account all 
of the observational and analysis biases (associated with both X-ray and SZ effect), which 
is beyond the scope of this paper.  We expect that it will be possible to get a much more 
firm handle on these differences in the near future.  First, the number of high redshift 
clusters observed through the SZ effect is increasingly rapidly (see Reese et al. 2002) 
and with it comes improved statistics.  Second, and perhaps more important, a number of 
new experiments (or substantial upgrades to existing ones) are already under construction 
and will greatly improve the quality of the SZ effect observations.  For example, the 
bandwidths of the future {\it SZA} and the upgraded {\it OVRO} array are expected to be 
nearly an order of magnitude larger than that of current interferometers and, as such, 
will lead to substantial improvements in the signal-to-noise ratios of SZ effect data.  
Tighter constraints on the SZ effect surface brightness profiles of clusters will then be 
possible and, in turn, more stringent limits can be placed on the entropy distributions 
of these clusters.  Indeed, we demonstrate below that observations with the {\it SZA} and 
the upgraded {\it OVRO} 
{\epsscale{1.0}
\plotone{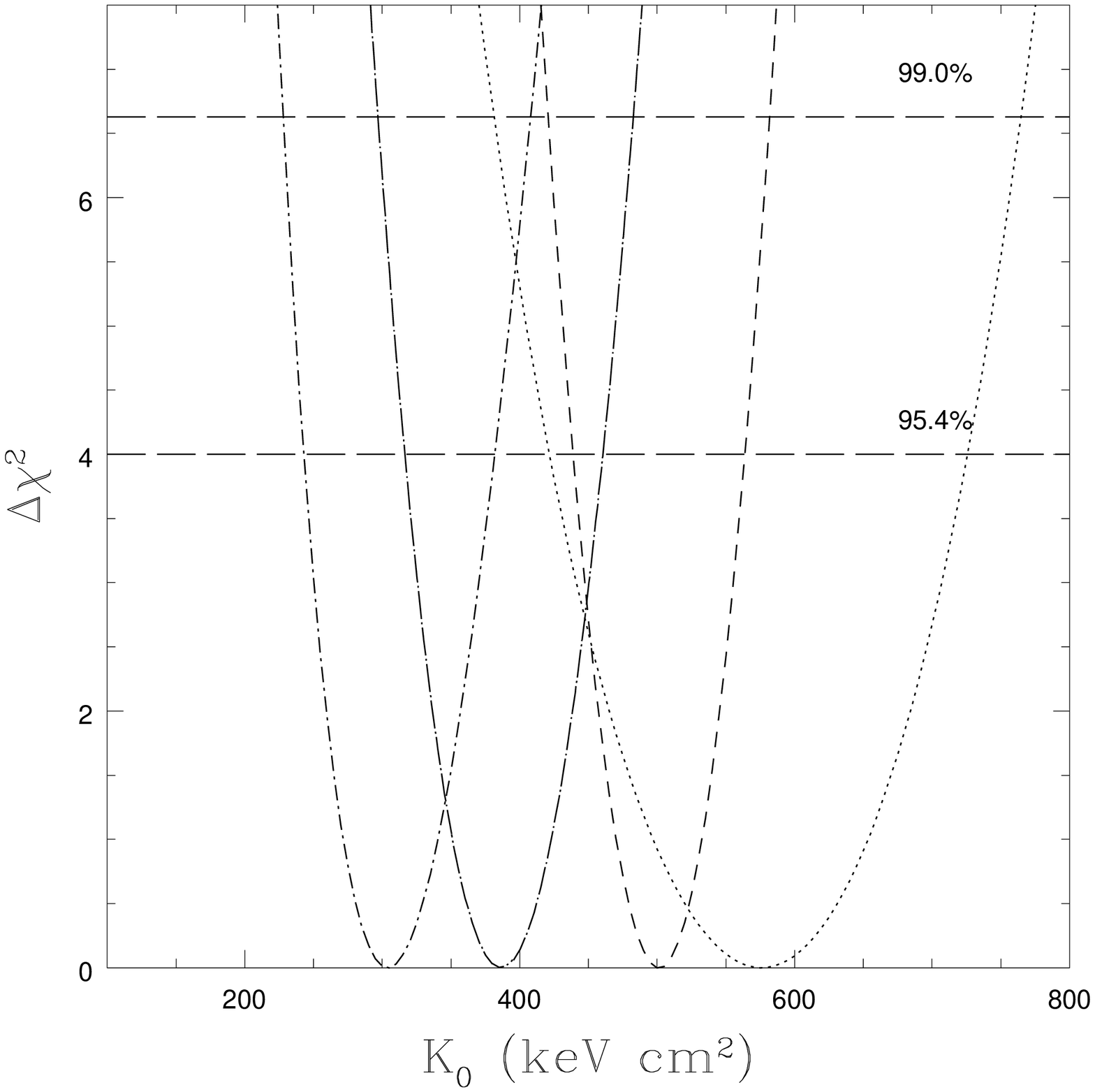}
{Fig. 8. \footnotesize
Constraints on $K_0$.  The dot-short-dashed, dot-long-dashed, dashed, and
dotted lines represent the $y_0-T_X$, $S_{\nu,arc}/f_{\nu}-L_X$, $y_0-M(r_{500})$, and
$S_{\nu,arc}/f_{\nu}-y_0$ relations (the four most entropy floor-sensitive relations),
respectively.  The cooling flow clusters A1835 and RXJ1347.5-1145 have been excluded
from the fits.
}}
\vskip0.1in
\noindent
array will allow one to constrain the amount of excess entropy in 
clusters all the way out to $z \sim 2$ and, best of all, without the need for any X-ray 
results. 
	
\section{The Future: Observations with SZA/OVRO}

It was briefly discussed in \S 3.1 that current SZ effect data alone cannot tightly 
constrain the three parameters of the surface brightness $\beta$ model.  A large degeneracy 
is present in the shape parameters, $\beta$ and $\theta_c$ (e.g., Carlstrom et al. 1996; 
Grego et al. 2000; 2001).  A number of authors have dealt with this problem by using the 
values of $\beta$ and $\theta_c$ obtained from modeling the {\it X-ray} surface brightness 
of the cluster (e.g., Patel et al. 2000; Grainge et al. 2002a; 2002b; LaRoque et al. 
2003).  
In the case of low/intermediate redshift clusters, X-ray observations provide tighter 
constraints on these parameters than do SZ effect observations.  However, a truly 
X-ray-independent probe of the ICM requires that only SZ effect data be used in the 
analysis.  By using a joint maximum-likelihood fit to both the X-ray and SZ effect surface 
brightnesses of clusters, Reese et al. (2000; 2002) found a compromise between these two 
scenarios.  Pushing the analysis of the ICM to higher redshifts than considered by Reese et 
al. (2000; 2002), however, will likely bring us into a regime that is uniquely accessible 
to SZ effect observations.  Fortunately, substantial upgrades are planned for currently 
operational arrays [for example, the current {\it BIMA} and {\it OVRO} arrays are being 
upgraded and merged into the Combined Array for Research in Millimeter-wave Astronomy 
({\it CARMA})$^6$] and a number of new interferometers are planned as well (e.g., {\it 
SZA}$^7$, {\it AMiBA}, and {\it AMI}; Holder et al. 2000; Lo et al. 2000; Kneissl et al. 
2001).  These experiments will substantially improve the quality of SZ effect 
observations.  Below, we compare the expected performance of the {\it SZA} and the (soon 
to be) upgraded {\it OVRO} array with the currently operational {\it BIMA/OVRO} arrays.  We 
show that the degeneracy in the $\beta-\theta_c$ plane is greatly reduced with data 
from the {\it SZA} or the upgraded {\it OVRO} array and this will allow for accurate 
determinations of $y_0$ and $S_{\nu}$ and without the need for any X-ray results.  

\footnotetext[6]{For information on the {\it CARMA} see
http://www.mmarray.org}

\footnotetext[7]{For information on the {\it SZA} see
http://astro.uchicago.edu/sze/survey.html}

\subsection{Mock Observations}

To compare the arrays, we first generate ``mock'' observations of model galaxy clusters.  We 
follow a method similar to that outlined by Holder et al. (2000).  First, we create a 
Compton parameter map, $y(\theta_x,\theta_y)$, for each model cluster.  These maps are 
converted into SZ effect intensity decrement maps.  Before we can do this, however, we must 
assume an observing frequency for each of the interferometers.  The current {\it BIMA} and 
{\it OVRO} arrays use amplifiers which operate over the range $26-36$ GHz (Reese et al. 
2000).  We assume a frequency centered on 30 GHz for these arrays.  The upgraded {\it OVRO} 
array and the {\it SZA} are expected to have amplifiers which operate at $26-36$ GHz and at 
$85-115$ GHz.  We assume frequencies centered on 30 and 90 GHz for these arrays.  Thus, we 
generate six different decrement maps for each model cluster: 30 GHz maps for the current 
{\it BIMA/OVRO} arrays and 30 and 90 GHz maps for the {\it SZA} and upgraded {\it OVRO} 
array.

Interferometers do not image the sky.  Rather, they measure the Fourier transform of the SZ 
effect intensity decrement multiplied by the primary beam.  A particular value of this 
observable is referred to as a ``visibility'' and is given by

\begin{equation}
V(u,v) = j_{\nu}\int y(\theta_x,\theta_y) A(\theta_x,\theta_y) e^{2 \pi i (u \theta_x + 
v \theta_y)} d\theta_x d\theta_y
\end{equation}

\noindent where $u$ and $v$ are the conjugate variables, $\theta_x$ and $\theta_y$ are the 
projected sky coordinates, and $A(\theta_x,\theta_y)$ is the primary beam.  At 30 GHz, the 
sensitivity patterns for the primary beams of the {\it BIMA} and {\it OVRO} arrays (both 
current and upgraded) and the {\it SZA} are, or will be, well-represented by Gaussians with 
FWHM $\approx$ $6.6\arcmin$, $4.2\arcmin$, and $10-12\arcmin$, respectively.  We assume a 
FWHM = $10.8\arcmin$ for the SZA.  For more on the above formalism see White et al. (1999).

We multiply the SZ effect maps by the Gaussian primary beams described above and Fourier 
transform the result.  We discard all visibilities for baselines ($R_{u,v} \equiv 
\sqrt{u^2+v^2}$) that are not probed by these interferometers.  At 30 GHz, the {\it BIMA} 
and {\it OVRO} arrays probe multipole moments of $\ell = 2 \pi R_{u,v}$ $\gtrsim 4000$ and 
$7000$, respectively.  We assume uniform coverage over the ranges $4200 \lesssim \ell 
\lesssim 20000$ and $7200 \lesssim \ell \lesssim 20000$ for these arrays, respectively.  
The smaller 3.5 m dishes of the {\it SZA} will be able to probe shorter baselines (larger 
angular scales) and are expected to sample $\ell \gtrsim 2000$.  We assume uniform coverage 
over the range $2400 \lesssim \ell \lesssim 20000$ (also at 30 GHz).  

Finally, we add random Gaussian noise to each of the mock observations.  At 30 GHz, we 
assume system temperatures (scaled to the atmosphere) of 45 K for the {\it BIMA} and {\it 
OVRO} arrays and 30 K for the {\it SZA}.  At 90 GHz, we assume system temperatures of 100 K 
for the {\it SZA} and for the upgraded {\it OVRO} array. The current {\it BIMA/OVRO} arrays 
have effective noise bandwidths of approximately 550/1000 MHz, while both the upgraded {\it 
OVRO} array and the SZA are expected to have bandwidths of 8 GHz.  A total integration time 
of 40 hours is assumed for each of the arrays at both 30 and 90 GHz.  For specificity, 
the mock observations for the current {\it BIMA} and {\it OVRO} arrays have rms noise 
levels of approximately 150 $\mu$Jy beam$^{-1}$ and 70 $\mu$Jy beam$^{-1}$, respectively, 
while the upgraded {\it OVRO} array has noise levels of 25 $\mu$Jy beam$^{-1}$ and 50 
$\mu$Jy beam$^{-1}$ at 30 and 90 GHz (respectively) and the {\it SZA} has noise levels of 
40 $\mu$Jy beam$^{-1}$ and 140 $\mu$Jy beam$^{-1}$ at 30 and 90 GHz, respectively. 

\subsection{Analysing the Mock Observations}

To analyse the mock observations, we follow the method described in a number of 
observational papers based on genuine {\it BIMA/OVRO} data (e.g., Carlstrom et al. 1996; 
Grego et al. 
2000; 2001; Reese et al. 2000; 2002; Joy et al. 2001; LaRoque et al. 2003).  We model the 
observations with the isothermal $\beta$ models$^8$.  We create model SZ effect maps for   
for various choices of the three free parameters ($y_0$, $\theta_c$, and $\beta$), multiply 
the maps by the appropriate primary beams, Fourier transform the maps, and compare the 
results to the mock observations in \S4.1 via a $\chi^2$ statistic.  The best-fit parameters 
are those which result in the minimum value of $\chi^2$.  Note that the comparisons between 
the $\beta$ models and the mock observations are done in the Fourier domain (also referred 
to as the $u-v$ plane), where the noise characteristics and spatial filtering of the 
interferometers are well understood.  Like the observers, we do not ``deconvolve'' the mock 
observations for analysis.

\footnotetext[8]{Note. --- we do not assume that the model cluster used to make the mock 
observations is necessarily isothermal.}

To demonstrate the quality of the data to be produced by the {\it SZA} and the upgraded {\it 
OVRO} array, we plot in Figure 9 confidence contours for the $\beta$ and $\theta_c$ 
parameters from ``observations'' of a $T_X = 6.7$ keV ($M_{tot} \approx 5.6 \times 10^{14} 
M_{\odot}$) cluster at $z = 1$.  The third free parameter, $y_0$, is allowed to assume its 
best-fit value at each pair of $\beta$ and $\theta_c$ (i.e., the plot is a projection of 
the confidence ``volume'' onto the $\beta-\theta_c$ plane).  The contours correspond 
to a $\Delta \chi^2$ and the filled squares indicate the best-fit models.  The projection 
of $\Delta \chi^2 = 1$ line onto the axes gives the 68\% confidence interval for each of 
the two parameters.  The contour labeled ``current'' is the result of fitting the 
SZ effect surface brightness model to the mock observations with the currently 
operational {\it BIMA/OVRO} arrays (both arrays have similar sensitivity patterns).  The 
contours labeled ``SZA'' and ``OVRO'' are the results of simultaneously fitting the 
surface brightness model to the mock 30 and 90 GHz {\it SZA} observations and the mock 30 
and 90 GHz upgraded {\it OVRO} observations, respectively.  Lastly, the contour labeled 
``SZA+OVRO'' is the result of simultaneously fitting all four upgraded OVRO and SZA mock 
observations. 

First, it is obvious that the ``current'' mock observational data do not tightly constrain 
$\beta$ and $\theta_c$ and, furthermore, these parameters are strongly correlated, as found 
when modeling genuine {\it BIMA/OVRO} data (see, e.g., Fig. 3 of Grego et al. 2000; Fig. 2 
of Grego et al. 2001).  This 
{\epsscale{1.0}
\plotone{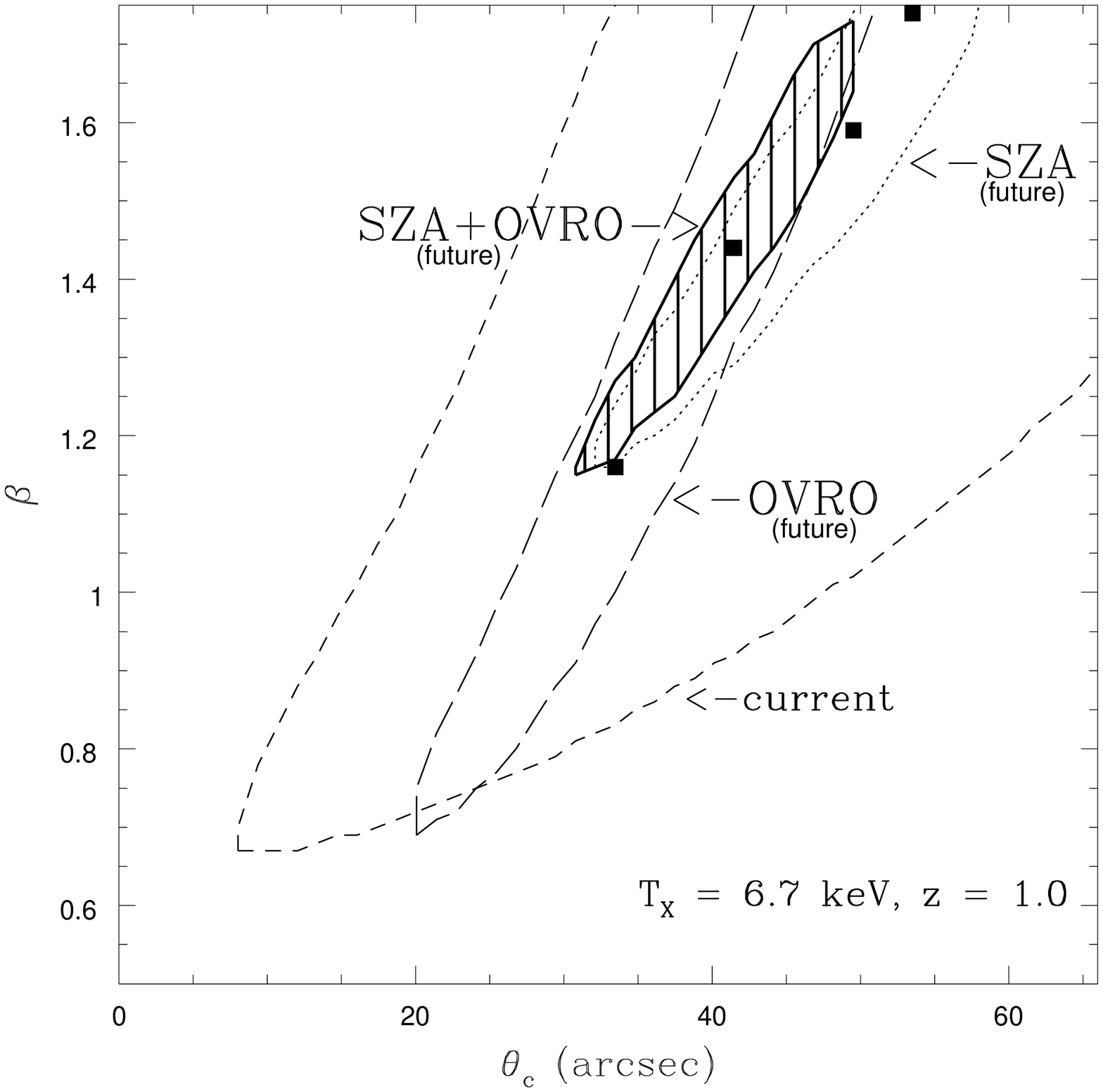}
{Fig. 9. \footnotesize
Comparison of constraints on the surface brightness profile of a distant,
massive cluster.  The filled squares indicate the best-fit parameters from modeling
``mock'' observations of the model cluster (the square at $\beta = 1.74$ is for
``current''
data while the square at $\beta = 1.59$ is for ``SZA'' data).  The contours give the 68\%
confidence regions on the two parameters ($\Delta \chi^2 = 1$).  The normalization, $y_0$,
is allowed to assume its best-fit value at each point in the plot (see text).
}}
\vskip0.1in
\noindent
gives us confidence that we have produced realistic mock 
observations.  The large degeneracy in the $\beta-\theta_c$ plane is why a number authors 
elect to use X-ray constraints on these parameters instead or, in the case of Reese et al. 
(2000; 2002), use a simultaneous fit to both X-ray and SZ effect data.  However, data from 
the upcoming {\it SZA} and the upgraded {\it OVRO} array will be able to place much tighter 
constraints on these parameters (compare the ``OVRO'' and ``SZA'' contours with the 
``current'' contour).  Note that the correlation between $\beta$ and $\theta_c$ remains for 
the {\it SZA} and {\it OVRO} ``data'', but its size has been dramatically reduced.  A 
simultaneous fit to both the {\it SZA} and upgraded {\it OVRO} data yields even better 
constraints on these parameters (shaded region).

To get an idea of how well the future SZ effect data can constrain $K_0$, the entropy floor 
level, we use the 68\% confidence volume ($\beta$, $\theta_c$, $y_0$) for the ``SZA+OVRO'' 
contour to ``measure'' $y_0$ and $S_{\nu,arc}/f_{\nu}$.  The inferred statistical 
uncertainty associated with the central Compton parameter and integrated SZ effect flux 
density within the central $1\arcmin$ for this cluster is only about 10\% and 15\%, 
respectively.  Comparing this to the predicted $S_{\nu,arc}/f_{\nu}-y_0$ relations (Figure 
2 of MBHB03), it should, therefore, be possible to constrain the entropy floor level of 
this cluster to within 50-75 keV cm$^2$ or so.  This is comparable with statistical 
uncertainties associated with X-ray measurements of nearby clusters.  This is 
remarkable considering that no X-ray ``data'' was used in the analysis and the cluster lies 
at $z = 1$.  We also find that reasonably accurate measurements of $K_0$ are possible for 
clusters all the way out to $z \sim 2$.

The {\it SZA} and upgraded {\it OVRO} array will be excellent tools for probing the 
non-gravitational entropy of distant clusters.  Because X-ray data will not be required to 
constrain the shapes of the SZ effect surface brightness profiles of clusters observed with 
these planned interferometers, comparisons of data to predicted scalings (such as 
$S_{\nu,arc}/f_{\nu}-y_0$) will provide {\it independent} constraints on the properties of 
the intracluster gas.  It will then also be possible to take advantage of the 
redshift-independence of the SZ effect and monitor the evolution of non-gravitational 
processes in clusters right back to the epoch of cluster formation itself.     

\section{Discussion}

Up until now, measurements (both direct and indirect) of the entropy floors of massive 
galaxy clusters have been limited to X-ray observations.  Furthermore, these past 
X-ray studies have generally focused on nearby ($z \sim 0$) clusters and, as such, 
little is known about the evolution of the entropy floor (and the non-gravitational 
processes that produce it) with cosmic time.  In the companion paper, we explored the 
extent to which the thermal Sunyaev-Zeldovich effect is modified by the presence of an 
entropy floor.  Because it depends differently on the temperature and density of ICM and, 
also, because it is redshift-independent, the SZ effect could potentially be a very 
powerful, independent test of the entropy floors of even the most distant galaxy 
clusters.  

The central focus of the present paper was to compare our theoretical relations from 
MBHB03 (including one that can potentially be measured through SZ effect observations 
only) to available high redshift SZ effect data from the literature to determine if the 
SZ effect data support the presence of an entropy floor and, if so, how does the inferred 
level of that floor compare with that required to match local X-ray trends.  This is the 
first time such a comparison has been done and we have made use of the largest compilation 
of high $z$ SZ effect clusters to date.  A detailed analysis of seven different SZ effect 
scaling relations indicates that the entropy floor in clusters with $0.14 \lesssim z 
\lesssim 0.78$ is between $300$ keV cm$^2$ $\lesssim K_0 \lesssim 600$ keV cm$^2$ and there 
are no strong indications for evolution in $K_0$ over that redshift interval.  Our estimate 
for the value of $K_0$ is remarkably similar to that derived from studies of X-ray scaling 
relations of nearby ($z \sim 0$) massive clusters, which suggest that $K_0 \gtrsim 300$ keV 
cm$^2$ (e.g., Tozzi \& Norman 2001; Babul et al. 2002; McCarthy et al. 2002).

At present, the source (or sources) of the ``excess'' entropy is still not known.  What 
constraints can be placed on the possible sources by the results of the present study?  
First, as in previous studies of X-ray scaling relations, our analysis indicates 
that the entropy of the ICM has been significantly raised by some non-gravitational 
process(es).  In terms of thermal energy, this corresponds to a few keV per particle for 
massive clusters.  This means that supernovae explosions probably cannot be the sole 
contributor to the entropy floor, since they are expected to impart $\lesssim 0.3$ keV 
per particle (e.g., Valageas \& Silk 1999; Balogh et al. 1999; Wu et al. 2000).  This was 
previously known but the present study, which offers an independent examination of the 
ICM, reinforces this conclusion.  Recently, it has been speculated that quasar outflows 
could be source of the excess entropy (e.g., Nath \& Roychowdhury 2002).  The entropy 
requirements deduced in the present study (and previous X-ray studies) are probably met 
by quasar outflows but it isn't yet known what mechanism (if any) couples the outflows to 
the ambient ICM.  Alternatively, and somewhat paradoxically, radiative cooling has also 
been shown to raise the mean entropy of ICM.  It is possible that cooling in combination 
with supernovae and/or quasar outflows could be responsible for the observed SZ effect 
and SZ effect-X-ray relations.  Whatever the source may be, it must reproduce the fact 
that $K_0$ does not change significantly out to $z \sim 0.7$.  It could well be that this 
trend will become a critical piece of information for discriminating between the various 
theoretical models currently being proposed.  An unchanging value of $K_0$ with redshift  
is obviously consistent with the generic ``preheating'' scenario; however, it remains to 
be seen whether it is consistent with more realistic heating models that distribute 
entropy non-uniformly and over an extended periods of time.  Without a detailed 
analysis, it is difficult to say whether or not it is consistent with radiative cooling 
contributing significantly to the excess entropy.  We are currently in the process of 
examining the effects of radiative cooling on SZ effect scaling relations (c.f. 
the discussion in the companion paper, MBHB03).

Current SZ effect data cannot tightly constrain the surface brightness profiles 
of clusters.  This prevents the SZ effect from being used as an independent (of X-ray) 
probe of the entropy floor and the ICM in general.  Thus, any advantages that the SZ effect 
has over the X-ray emission (e.g., redshift independence) are severely diminished because 
X-ray data is needed to help constrain the shape of the surface brightness profiles. 
Therefore, an additional aim was to examine the ability of the next generation of SZ 
effect experiments to probe non-gravitational entropy in distant clusters.  
We have shown that the {\it SZA} and the upgraded {\it OVRO} array will produce high 
signal-to-noise data that will allow one to tightly constrain the surface brightness 
profiles of even very distant clusters and without the need for any X-ray results.  As in 
the present study, these surface brightness profiles can then be compared to theoretical 
predictions in order to place stringent constraints on the level of the entropy floor. It 
will be very interesting to see if the trend of constant $K_0$ (with redshift) deduced 
here holds up and, if so, to determine how far back in redshift it extends.  

Aside from the appearance and structure of individual clusters, our work has implications 
for universal SZ effect quantities, such as the SZ effect angular power spectrum, SZ effect 
cluster source counts, and the mean Compton parameter of the universe.  These quantities can 
be used to measure cosmological parameters and test cluster formation scenarios (see 
Carlstrom, Holder, \& Reese 2002 for a comprehensive review).  A number of studies have 
already examined how non-gravitational gas physics modifies these quantities (e.g., Holder 
\& Carlstrom 2001; Cavaliere \& Menci 2001; da Silva et al. 2001; Springel et al. 2001; 
White et al. 2002).  However, none of these studies have invoked entropy 
injection at the high level estimated in the present analysis.  Generally, low levels of 
entropy injection ($K_0 \sim 100$ keV cm$^2$), which are consistent with X-ray 
measurements from low mass groups (Ponman et al. 1999), were implemented.  We are in the 
process of investigating how the power spectrum, source counts, and mean Compton 
parameter are modified by higher initial entropies.

\vskip0.25in

We thank Erik Reese for making available his SZ effect data prior to publication.  We also 
thank Ann Gower, John Carlstrom, Peng Oh, Kathy Romer, and Mark Voit for helpful 
discussions.  A. B. would like to acknowledge the hospitality extended to him by the 
Canadian Institute for Theoretical Astrophysics during his tenure as CITA Senior Fellow.  
I. G. M. is supported by a postgraduate scholarship from the Natural Sciences and 
Engineering Research Council of Canada (NSERC).  A. B. is supported by an NSERC operating 
grant, G. P. H. is supported by the W. M. Keck Foundation, and M. L. B. is supported by a 
PPARC rolling grant for extragalactic astronomy and cosmology at the University of 
Durham.

\clearpage

\begin{deluxetable}{lllll}
\tablecaption{SZ effect observational data
\label{tab1}}
\tablewidth{25pc}
\tablehead{
\colhead{Cluster}                      &
\colhead{z} &
\colhead{$\log{y_0}$}        &
\colhead{$\log{S_{\nu,arc}/f_{\nu}}$} &
\colhead{Ref.}
}
\tablenotetext{~}{Note. --- $S_{\nu,arc}/f_{\nu}$ expressed in mJy.}
\tablenotetext{a}{Reese et al. (2002)}
\tablenotetext{b}{Jones et al. (2003)}
\tablenotetext{c}{Holzapfel (1996)}
\tablenotetext{d}{Holzapfel et al. (1997)}
\tablenotetext{e}{Lamarre et al. (1998)}
\tablenotetext{f}{Saunders et al. (1999)}
\tablenotetext{g}{Mauskopf et al. (2000)}
\tablenotetext{h}{Pointecouteau et al. (2002)}
\tablenotetext{i}{Pointecouteau et al. (2001)}
\tablenotetext{j}{Komatsu et al. (1999)}
\tablenotetext{k}{Grainge et al. (2002a)}
\tablenotetext{l}{Hughes \& Birkinshaw (1998)}
\tablenotetext{m}{Cotter et al. (2002)}
\startdata
A1413           &0.143 & $-3.794^{+0.053}_{-0.060}$ & $0.954_{-0.070}^{+0.062}$ & a (b) \\
A2204		&0.152	& $-3.744^{+0.111}_{-0.150}$ & \nodata			& c \\
A1914		&0.171	& $-3.798$ 		     & $0.934$                  & b \\
A2218           &0.176 & $-3.862^{+0.056}_{-0.081}$ & $0.922_{-0.086}^{+0.060}$ & a (b)\\
A665            &0.182 & $-3.864^{+0.081}_{-0.100}$ & $0.939_{-0.104}^{+0.084}$ & a\\
A1689           &0.183 & $-3.489^{+0.029}_{-0.027}$ & $1.169_{-0.036}^{+0.038}$ & a\\
A520            &0.199 & $-3.906^{+0.058}_{-0.067}$ & $0.914_{-0.075}^{+0.065}$ & a\\
A2163           &0.203 & $-3.448^{+0.031}_{-0.033}$ & $1.363_{-0.036}^{+0.034}$ & a (d,e)\\
A773            &0.217 & $-3.626^{+0.052}_{-0.059}$ & $1.132_{-0.103}^{+0.080}$ & a (b,f)\\
A2261           &0.224 & $-3.497^{+0.048}_{-0.054}$ & $1.137_{-0.087}^{+0.085}$ & a\\
A1835           &0.253 & $-3.328^{+0.029}_{-0.027}$ & $1.146_{-0.050}^{+0.048}$ & a (g)\\
A697            &0.282 & $-3.577^{+0.052}_{-0.052}$ & $1.179_{-0.091}^{+0.082}$ & a (b)\\
A611            &0.288 & $-3.795^{+0.066}_{-0.066}$ & $0.807_{-0.179}^{+0.141}$ & a\\
Zwicky 3146     &0.291 & $-3.792^{+0.066}_{-0.077}$ & \nodata			& c\\
A1995           &0.319 & $-3.717^{+0.032}_{-0.037}$ & $0.943_{-0.125}^{+0.092}$ & a\\
MS1358.4+6245   &0.327 & $-3.832^{+0.047}_{-0.053}$ & $0.720_{-0.090}^{+0.083}$ & a\\
A370            &0.375 & $-3.628^{+0.154}_{-0.083}$ & $1.145_{-0.180}^{+0.204}$ & a\\
RXJ2228+2037    &0.421 & $-3.620$		     & $1.024$                  & h\\
RXJ1347.5-1145  &0.451 & $-3.130^{+0.037}_{-0.040}$ & $1.245_{-0.072}^{+0.069}$ & a (i,j)\\
Cl0016+16       &0.546 & $-3.632^{+0.035}_{-0.038}$ & $1.056_{-0.059}^{+0.054}$ & a (k,l)\\
MS0451.6-0305   &0.550 & $-3.571^{+0.027}_{-0.031}$ & $1.041_{-0.091}^{+0.076}$ & a\\
MS1137.5+6625   &0.784 & $-3.814^{+0.048}_{-0.055}$ & $0.608_{-0.365}^{+0.243}$ & a (m)\\
\enddata
\end{deluxetable}

\begin{deluxetable}{llllll}
\tablecaption{X-ray observational data
\label{tab2}}
\tablewidth{30pc}
\tablehead{
\colhead{Cluster}                      &
\colhead{$\log{M(r_{500})}$} &
\colhead{$\log{T_X}$} &
\colhead{$\log{L_X}$} &
\colhead{Type} &
\colhead{Ref.}
}
\tablenotetext{~}{Note. --- Masses, temperatures, and bolometric luminosities are expressed 
in $M_{\odot}$, keV and ergs s$^{-1}$, respectively.}
\tablenotetext{a}{Ettori \& Fabian (1999)}
\tablenotetext{b}{Allen (2000)}
\tablenotetext{c}{White, Jones, \& Forman (1997)}
\tablenotetext{d}{Jones et al. (2003)}
\tablenotetext{e}{Machacek et al. (2002)}
\tablenotetext{f}{Markevitch \& Vikhlinin (2001)}
\tablenotetext{g}{Mushotzky \& Scharf (1997)}
\tablenotetext{h}{Allen \& Fabian (1998)}
\tablenotetext{i}{White (2000)}
\tablenotetext{j}{Novicki, Sornig, \& Henry (2002)}
\tablenotetext{k}{Vikhlinin et al. (2002)}
\startdata
A1413           & 14.86 & $0.929^{+0.062}_{-0.043}$& $45.082^{+0.018}_{-0.018}$ & CF & a,b,c 	\\
A2204           & 14.88 & $0.964^{+0.104}_{-0.055}$& $45.407$ & CF & a,b,c	\\
A1914           & \nodata & $0.934^{+0.018}_{-0.020}$& \nodata & NCF & d	\\
A2218           & 14.73 & $0.839^{+0.030}_{-0.033}$& $44.974^{+0.018}_{-0.018}$ & NCF & a,e,c	\\
A665            & 14.83 & $0.944^{+0.042}_{-0.047}$& $45.124^{+0.013}_{-0.013}$ & NCF & a,f,c	\\
A1689           & 15.06 & $1.000^{+0.049}_{-0.036}$& $45.423^{+0.010}_{-0.010}$ & CF & a,b,c	\\
A520            & 14.75 & $0.921^{+0.038}_{-0.036}$& $45.070$ & NCF & a,b,c	\\
A2163           & 15.11 & $1.090^{+0.044}_{-0.041}$& $45.732$ & NCF & a,f,c	\\
A773            & \nodata & $0.968^{+0.016}_{-0.016}$& $45.278$ & NCF & b,g	\\
A2261           & \nodata & $1.037^{+0.188}_{-0.098}$& $45.447$ & CF & b,h	\\
A1835           & \nodata & $0.991^{+0.092}_{-0.062}$& $45.725$ & CF & b,g	\\
A697            & \nodata & $1.009^{+0.074}_{-0.090}$& \nodata & NCF & i	\\
A611            & \nodata & $0.836^{+0.029}_{-0.030}$& \nodata & CF & i	\\
Zwicky 3146     & 14.91 & $1.053^{+0.180}_{-0.119}$&  $45.525$ & CF & a,b,g	\\
A1995           & \nodata & $1.016^{+0.053}_{-0.053}$& $45.241^{+0.010}_{-0.010}$ & CF & j,j	\\
MS1358.4+6245   & 14.81 & $0.875^{+0.289}_{-0.097}$& $45.093^{+0.014}_{-0.014}$ & CF & a,b,j	\\
A370            & \nodata & $0.848^{+0.031}_{-0.032}$& $45.236^{+0.010}_{-0.010}$ & NCF & i,j	\\
RXJ2228+2037    & \nodata & \nodata& \nodata & NCF & \nodata	\\
RXJ1347.5-1145  & \nodata & $1.422^{+0.112}_{-0.272}$& $46.093^{+0.003}_{-0.003}$ & CF & b,j	\\
Cl0016+16       & \nodata & $0.911^{+0.041}_{-0.045}$& \nodata & NCF & i	\\
MS0451.6-0305   & \nodata & $0.908^{+0.041}_{-0.045}$& $45.610$ & NCF & k,k	\\
MS1137.5+6625   & \nodata & $0.799^{+0.027}_{-0.028}$& $45.158$ & NCF & k,k	\\
\enddata
\end{deluxetable}

\end{document}